\documentclass[fleqn,usenatbib]{mnras}

\usepackage{newtxtext,newtxmath}

\usepackage[T1]{fontenc}

\DeclareRobustCommand{\VAN}[3]{#2}
\let\VANthebibliography\thebibliography
\def\thebibliography{\DeclareRobustCommand{\VAN}[3]{##3}\VANthebibliography}

\usepackage{graphicx}	
\graphicspath{{figures/}}	
\usepackage{amsmath}	
\usepackage{amsfonts}
\usepackage{bm}		

\title[Analysis of accretion disc structure and stability]{Analysis of accretion disc structure and stability using open code for vertical structure}

\author[A. S. Tavleev et al.]{A. S. Tavleev,$^{1,2}$\thanks{E-mail: tavleev@astro.uni-tuebingen.de} G. V. Lipunova,$^{1,3}$ and K. L. Malanchev$^{1,4}$ \\
$^{1}$ Sternberg Astronomical Institute, Moscow M. V. Lomonosov State University, 13 Universitetski pr., 119234, Moscow, Russia \\
$^{2}$ Institut f\"ur Astronomie und Astrophysik, Kepler Center for Astro and Particle Physics, Universit\"at T\"ubingen, Sand 1, 72076 T\"ubingen, Germany \\
$^{3}$ Max-Planck-Institut f\"ur Radioastronomie, Auf dem H\"ugel 69, 53121 Bonn, Germany \\
$^{4}$ Department of Astronomy, University of Illinois at Urbana-Champaign, 1002 W. Green St., IL 61801, USA
}

\date{Accepted 2023 June 19. Received 2023 June 15; in original form 2022 December 6
}

\pubyear{2023}

\begin{document}
\label{firstpage}
\pagerange{\pageref{firstpage}--\pageref{lastpage}}
\maketitle

\begin{abstract}
Radial structure of accretion discs around compact objects is often described using analytic approximations which are derived from averaging or integrating vertical structure equations. For non-solar chemical composition, partial ionization, or for supermassive black holes, this approach is not accurate. Additionally, radial extension of `analytically-described' disc zones is not evident in many cases. 
We calculate vertical structure of accretion discs around compact objects, with and without external irradiation, with radiative and convective energy transport taken into account. For this, we introduce a new open Python code, allowing different equations of state (EoS) and opacity laws, including tabular values. As a result, radial structure and stability `S-curves' are calculated for specific disc parameters and chemical composition. In particular, based on more accurate power-law approximations for opacity in the disc, we supply new analytic formulas for the farthest regions of the hot disc around stellar-mass object.
On calculating vertical structure of a self-irradiated disc, we calculate a self-consistent value of the irradiation parameter $C_{\rm irr}$ for stationary $\alpha$-disc. We find that, for a fixed shape of the X-ray spectrum, $C_{\rm irr}$ depends weakly on the accretion rate but changes with radius, and the dependence is driven by the conditions in the photosphere and disc opening angle. The hot zone extent depends on the ratio between irradiating and intrinsic flux: corresponding relation for $T_{\rm irr,\, crit}$ is obtained.
\end{abstract}




\section{Introduction}

Disc accretion is a common astrophysical phenomenon widely observed thanks to high efficiency of energy conversion to emission. Brightest sources of the X-ray sky are explained by accretion of matter on compact objects, and visibility of such sources allows us to investigate physics operating there. Many X-ray sources are found in binary systems where the matter flows from one component to another and accretion discs are formed.

The standard model of viscous accretion discs~\citep{Shakura1972,ShakuraSunyaev1973} is based on the notion of the turbulent viscosity as a mechanism for the angular momentum transfer, allowing the matter, rotating around a central object, to move inwards and to emit gravitational energy converted to heat. Heat balance determines the vertical structure of the disc, that is, in the direction perpendicular to its symmetry plane. It is safe to assume that hydrostatic equilibrium holds in the vertical direction, meaning that the time to achieve the hydrostatic balance is shorter than other characteristic disc times. The thermal balance in the vertical direction occurs on a time-scale longer than a hydrostatic one, but faster than the disc evolves due to accretion rate variations. Thus, generally, the vertical disc structure could be studied separately from the radial one.

Considering accretion onto compact object of stellar masses, it is commonly assumed that the standard model describes well the disc regions that emit mainly in the optical. In this context, the details of the disc vertical structure are important since they determine the brightness and spectra of those regions.

It has been known for some time that the vertical structure of accretion disc is subject to various instabilities. Thermal-viscous instability is believed to be a cause of outbursts occurring rather periodically in some sources with accretion discs. A model based on the instability has been developed in a number of works (e.g., \citet{Hoshi1979, Smak1982_part1, Meyer1981, Meyer1982, Faulkner_etal1983_1, Faulkner_etal1983_2, Smak1984_part4}). Presently, it is referred to as the Disc Instability Model~\citep[or DIM,][]{Hameury1998,Lasota2001_review, Hameury2020_review}, see also \citet{Baginska_etal2021}. Details of DIM depend not only on the disc vertical structure at different radii but on the radial energy transport as well. To some extent, using the local analysis alone, it is possible to study the scenario with thermal-viscous instability on a basis of so-called S-curves~\citep{Meyer1981}, or equilibrium curves. An S-curve is a graphically depicted sequence of solutions of the vertical-structure equations, obtained at a single disc radius, in the coordinates of accretion rate or effective temperature versus the surface density (see Fig.~\ref{fig:S-curve} below). The positive slope of an S-curve represents the thermally and viscously stable state of the disc, whereas the negative slope represents the unstable state.

In X-ray transients, burst evolution depends crucially on the self-irradiation of the disc: heating by central X-rays can change the local state of the outer disc and, thus, the viscosity there~\citep{Tuchman_etal1990, Dubus_etal2001}. Vertical structure of a self-irradiated $\alpha$-disc has been calculated by \citet{Tuchman_etal1990, Dubus_etal1999}, who introduced a self-irradiation parameter. Self-consistent calculations of the vertical structure of irradiated discs have been preformed by \citet{Mescheryakov_etal2011} for fully ionized disc regions with opacity from the Opacity Project~\citep{Badnell_etal2005}.

In the current paper, we present results obtained with our new open \texttt{Python} code with modern values of opacity~\citep{Iglesias&Rogers1993, Iglesias&Rogers1996, Ferguson_etal2005} and equations of state~\citep{Rogers&Nayfonov2002}\footnote{Opacities of \citet{Iglesias&Rogers1993, Iglesias&Rogers1996} and \citet{Badnell_etal2005} differs by up to 10\% in area of typical disc parameters.}. The code calculates the vertical structure, S-curves, and radial profiles of optically thick accretion discs. We take into account X-ray irradiation by two methods and vertical transfer of energy by convection using an approach of the mixing-length theory. The code uses MESA package~\citep{Paxton_etal2011} for interpolation and sewing of the opacity and EoS tables.

For un-irradiated discs, we analyse physical conditions in the discs for a wide range of parameters. We examine to what extent analytical approximations for opacity laws, and, consequently, for radial dependencies, can be satisfactorily used. 

We analyse stability conditions for un-irradiated and self-irradiated discs. For self-irradiated discs, we also calculate the value of the self-irradiation parameter in the thermally stable disc parts, and analyse its dependence on the basic parameters of accretion disc. 

In section~\ref{sec:Modelling-of-accretion-disc-vertical-structure}, we present a system of equations of the vertical structure and boundary conditions, in particular, in the presence of external X-ray irradiation. The radial structure of the disc, resulting from solutions for vertical structure, is investigated in section~\ref{sec:Radial-structure}. The irradiation parameter is considered there as well. In section~\ref{sec:S-curves}, we construct and analyse S-curves. In section~\ref{sec:Discussion} we analyse and discuss the stability criterion of the irradiated disc. Summary is given in section~\ref{sec:Summary}. Appendix~\ref{appendix:Irradiation-formulas} reviews equations used in presence of external X-ray disc irradiation, while appendix~\ref{appendix:Brief-Code-description} contains a brief description of the code. Appendix~\ref{appendix:Vertical-structure} provides several examples of vertical structures calculated by the new code.

\section{Modelling of accretion disc vertical structure}
\label{sec:Modelling-of-accretion-disc-vertical-structure}

The vertical structure of $\alpha$-discs has been solved in a number of papers, including \citet{Smak1984_part4, Meyer1982, Hameury1998, Lasota_etal2008} where discs in X-ray transients were considered in particular. Note that they used previous values of opacity~\citep{Cox_etal1969, Cox_etal1976, Alexander1975} and EoS~\citep{Fontaine_etal1977}. \citet{KetsarisShakura1998, Suleimanov_etal2007, Malanchev_etal2017} have solved the vertical structure with analytical opacity coefficient and equation of state, which allows obtaining analytical radial structure (see Sect.~\ref{sec:Radial-structure}).

We use a cylindrical coordinate system ($r, \varphi, z$), where $z$ changes from $0$ in the symmetry plane to the semi-thickness of disc $z_0$ on the disc surface. We consider geometrically thin ($z_0 \ll r$) Keplerian ($\omega = \omega_{\rm K} = \sqrt{GM/r^3}$) stationary ($\partial / \partial t = 0$) optically thick ($\tau\gg1$) accretion disc.

\subsection{Basic equations}
\label{subsec:Basic-equations}

The vertical structure is described by the system of four ordinary differential equations \citep[see e.g.][]{shakura_etal2018}, which follows from the mass, energy and momentum conservation laws. For moderate accretion rates and small temperature gradients along the radius, the energy balance is local. First we consider discs without external heating by irradiation.
\begin{align}
    &\frac{{\rm d}P}{{\rm d}z} = -\rho\,\omega^2_{\rm K} z, \label{eq:P} \\ 
    &\frac{{\rm d}Q}{{\rm d}z} = \frac32 w_{r\varphi}\omega_{\rm K} = \frac32\omega_{\rm K} \alpha P, \label{eq:Q} \\
    &\frac{{\rm d}\ln T}{{\rm d}\ln P} \equiv \nabla = 
        \begin{cases}
        \nabla_{\rm rad}, \, \nabla_{\rm rad}\leq\nabla_{\rm ad}, \\
        \nabla_{\rm conv}, \, \nabla_{\rm rad}\geq\nabla_{\rm ad}, \\
        \end{cases} \label{eq:T} \\
    &\frac{{\rm d}\Sigma}{{\rm d}z} = -2\rho, \label{eq:Sigma} \\ 
    &z \in [0, z_0]. \nonumber
\end{align}
Here $P=P_{\rm rad} + P_{\rm gas}$ is the total pressure, $Q$ is the heating flux, which for an un-irradiated disc equals the viscous flux $Q_{\rm vis}$, and $T$ is the temperature. The mass coordinate $\Sigma(z)$ equals to zero at $z=0$ and the surface density of the disc $\Sigma_0$ at $z=z_0$. The last part of \eqref{eq:Q} includes the $\alpha$-prescription~\citep{ShakuraSunyaev1973}, where the absolute value of the $r\varphi$-component of tensor of viscous tensions $w_{r\varphi} = \alpha P$, and $\alpha$ is the turbulent parameter ($0<\alpha<1$).

If the energy is transported solely by radiation diffusion, equation~\eqref{eq:T} implies that
\begin{equation}
    \nabla = \nabla_{\rm rad} \equiv \frac{3\varkappa_{\rm R}}{4ac\omega^2_{\rm K} z} \frac{P}{T^4} Q,
    \label{eq:nabla_rad}
\end{equation}
where $\varkappa_{\rm R}$ is the Rosseland opacity coefficient, $a = 4\sigma_{\rm SB}/c$ is the radiation constant, $c$ is the speed of light. Otherwise, if $\nabla_{\rm rad}\geq\nabla_{\rm ad}$, the convective motions start to transfer energy according to the Schwarzschild~(\citeyear{Schwarzschild1958}) criterion. The corresponding temperature gradient $\nabla_{\rm conv}$ can be calculated according to the mixing length theory~\citep[see][]{Paczynski1969, Kippenhahn2012,Hameury1998}. Note that \citet{MalanchevShakura2015} considered viscous energy generation in convective cells and found that it made convection energy transfer less efficient. The thermodynamic values, needed to calculate the temperature gradient, for example $\nabla_{\rm ad}$, are obtained from the \texttt{eos}~module of the MESA code~\citep{Paxton_etal2011}.

In some cases, for the sake of comparison, we calculate the disc structure forcing
$\nabla = \nabla_{\rm rad}$ everywhere and call such models as `no convection'.

\subsection{Equation of state and opacity law}
\label{subsec:Equation-of-state-and-opacity-law}

System~(\ref{eq:P}--\ref{eq:Sigma}) should be supplemented by equation of state (EoS) and opacity law. They can be set both analytically or as tabular values. For analytical description, the ideal gas equation is adopted, while the opacity coefficient is approximated by a power-law function:
\begin{equation}
    \rho = \frac{\mu\,P_{\rm gas}}{\mathcal{R}\,T}, \qquad \varkappa_{\rm R} = \varkappa_0 \frac{\rho^{\zeta}}{T^{\gamma}}.
\end{equation}
Here $\mu$ is the molecular weight and $\varkappa_0$ is the dimension constant, which we give below is CGS units. For opacity, we consider the following options:
\begin{itemize}
    \item the Kramers law for bound-bound and free-free transitions with  $\zeta = 1, \gamma = 7/2$. For the solar chemical composition $\varkappa_0 = 5\cdot10^{24}$~\citep{FKR2002}.
    \item Two approximations by \citet{BellLin1994} to opacity produced by bound-free and free-free transitions with $\varkappa_0 = 1.5\cdot10^{20}, \zeta = 1, \gamma = 5/2$ and scattering off hydrogen atoms with $\varkappa_0 = 1\cdot10^{-36}, \zeta = 1/3, \gamma = -10$.
\end{itemize}
At $T\gtrsim 10^6$~K the opacity is dominated by the Thomson electron scattering, i.e. $\varkappa_{\rm R} = 0.34\,\rm cm^2 \, g^{-1}$.

\begin{figure}
    \center{\includegraphics[width=0.95\linewidth]{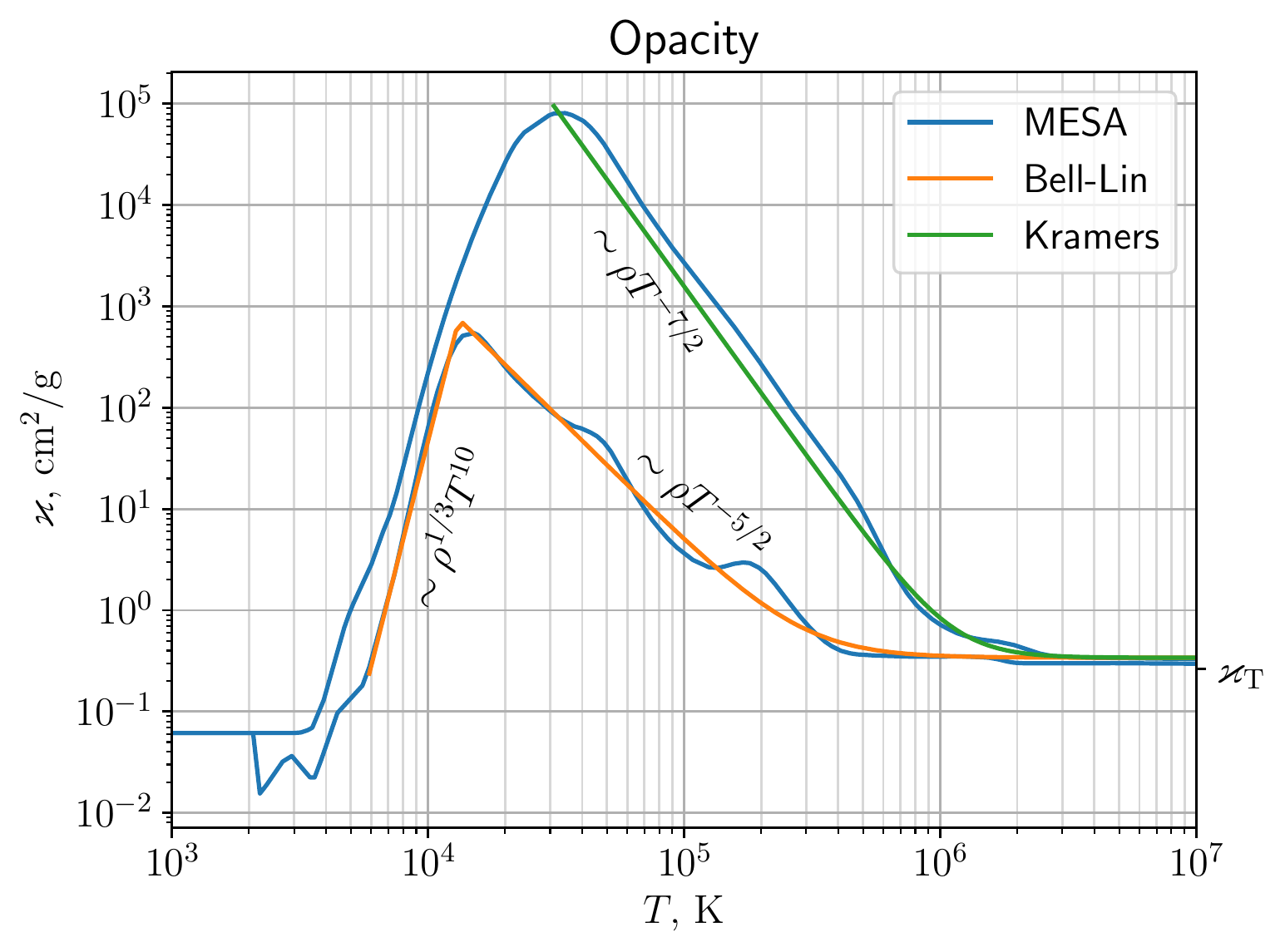}}
    \caption{Rosseland opacity coefficient, tabular and analytical, for solar chemical composition and for two bulk densities, $\rho = 10^{-4} {\rm~and~} 10^{-7}\, \rm g\, cm^{-3}$ (the upper and lower lines, respectively). The blue lines show interpolation of the tabular values (see Sect.~\ref{subsec:Equation-of-state-and-opacity-law}). Orange and green lines show Kramers~\citep{FKR2002} and \citet{BellLin1994} power-law approximations.}
    \label{fig:Opacity_all}
\end{figure}
\begin{figure*}
    \begin{minipage}{0.45\linewidth}
	    \center{\includegraphics[width=1.0\linewidth]{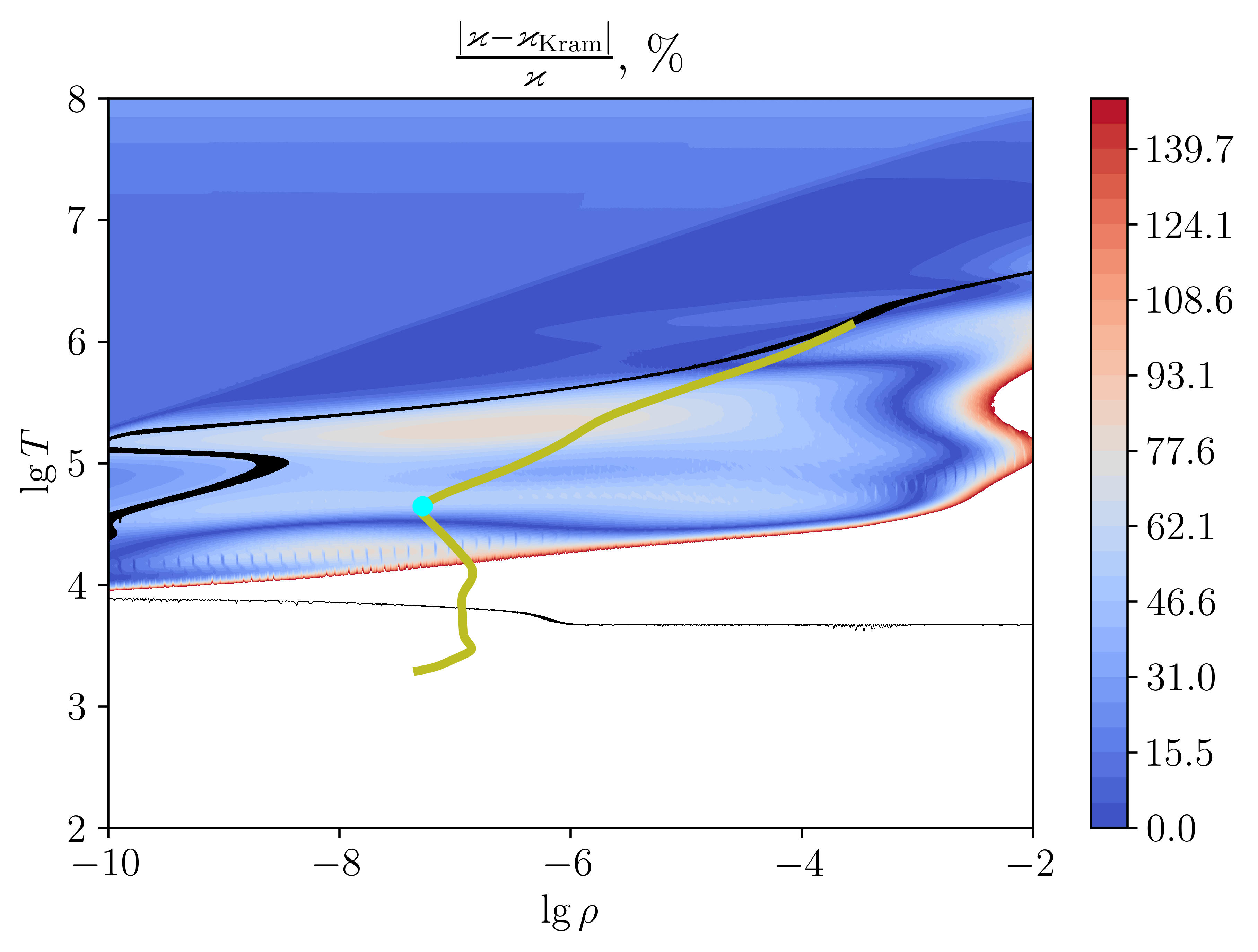}}
	\end{minipage}
	\begin{minipage}{0.45\linewidth}
	    \center{\includegraphics[width=1.0\linewidth]{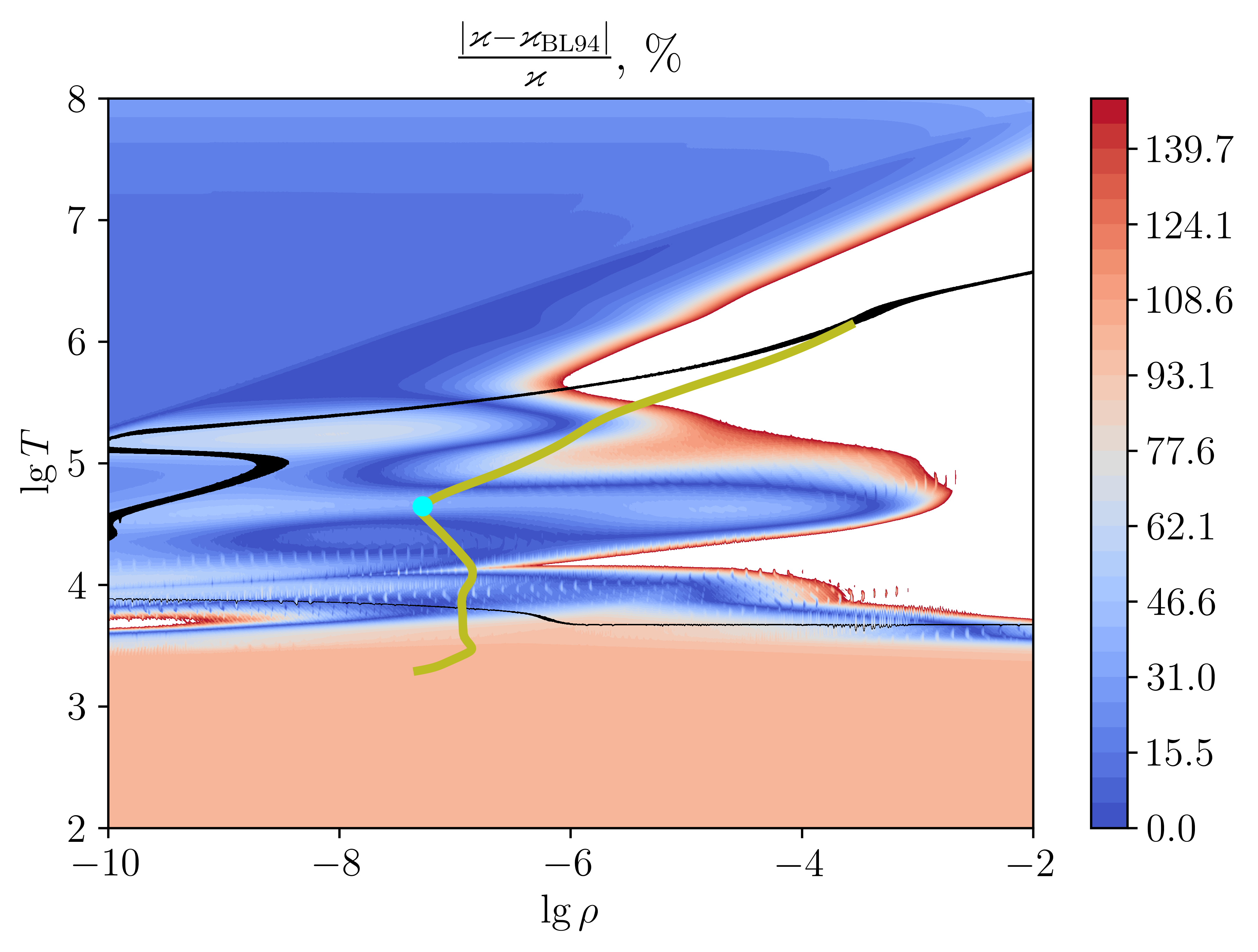}}
	\end{minipage}
	\caption{Shown are relative uncertainties between analytical approximations of opacity compared to tabular opacity values, obtained from MESA, as functions of density and temperature. Colour indicates the uncertainty in percents. In white regions the uncertainty is larger than 150\%. The left panel shows the uncertainty relative to the Kramers opacity law, and the right one, to the the approximations by \citet{BellLin1994} (see Sect.~\ref{subsec:Equation-of-state-and-opacity-law} for details). The black curve shows the range of $\rho$ and $T$, where the electron scattering $\varkappa_{\rm T}$ and absorption $\varkappa_{\rm ff}$ opacity coefficients are equal. The olive curve shows the $\rho-T$ relation for disc around $10\, M_{\sun}$ object, with turbulent parameter $\alpha = 0.1$ and accretion rate $\dot{M} = 10^{18} \rm g\, s^{-1}$, corresponding to Fig.~\ref{fig:r_plots_conv_ZS} below; the dot on the curve indicates the parameters at the outer hot-zone radius.}
	\label{fig:Opacity-diagram}
\end{figure*}

Tabular values of opacity~\citep{Iglesias&Rogers1993, Iglesias&Rogers1996, Ferguson_etal2005} and EoS~\citep{Rogers&Nayfonov2002} are obtained by interpolation using the \texttt{kap} and \texttt{eos} modules of the MESA code~\citep{Paxton_etal2011}. Their dependence on the temperature is shown in Fig.~\ref{fig:Opacity_all} for two values of density~$\rho$ along with the analytical approximations. \autoref{fig:Opacity-diagram} shows numerical differences between analytical and tabular opacities. The olive curve, superimposed on the diagram, represents the sequence of $\rho$ and $T$ obtained in our code for a disc around a $10\, M_{\sun }$ central star. The cyan dot in the middle of it marks a boundary between thermally stable and unstable disc zones. It can be concluded that the Kramers law approximates better the tabular opacity in the hotter parts of the disc, while the 'BL94' works better for the colder parts, where $T_{\rm eff} \sim (3-30)\times 10^3~$K.

\subsection{Boundary conditions}

We integrate system~(\ref{eq:P}--\ref{eq:Sigma}) starting from the surface of the disc, where $z=z_0$ and the optical depth $\tau=2/3$. The boundary conditions there are as follows:
\begin{align}
        Q(z_0) &= Q_0 \equiv Q_{\rm vis}(z_0) = \frac{3}{8\pi} \frac{F\omega_{\rm K}}{r^2}, \label{eq:Q_bound} \\
        T(z_0) &= T_{\rm eff} \equiv (Q_0/\sigma_{\rm SB})^{1/4}, \label{eq:T_bound} \\
        \Sigma(z_0) &= 0\, .\label{eq:Sigma_bound} 
\end{align}

Here $F=2\piup r^2 W_{r\varphi}$ is the viscous torque, and $W_{r_\varphi}\equiv\int_{-z_0}^{z_0}w_{r\varphi}{\rm d}z$ is the height-integrated viscous stress. In a stationary accretion disc the latter is derived from the angular momentum conservation and reads as follows: 
\begin{equation}
    F = \dot{M}h \left(1 - \sqrt{\frac{r_{\rm in}}{r}}\right) + F_{\rm in},
    \label{eq:Torque}
\end{equation}
where $h=\sqrt{GMr}$ is the specific angular momentum and $F_{\rm in}$ is the viscous torque at the inner radius $r_{\rm in}$~\citep[see e.g.,][]{shakura_etal2018}. Below we assume $W_{r_\varphi}(r_{\rm in})=0$. Notice that an arbitrary torque $F(r)$ can be set in the code.

To obtain the boundary condition for pressure, we write equation of hydrostatic equilibrium~(\ref{eq:P}) in the photosphere:
\begin{equation}
    \frac{{\rm d}P}{{\rm d}\tau} = \frac{\omega_{\rm K}^2 z}{\varkappa_{\rm R}}, \label{eq:Pph}
\end{equation} 
where the optical depth $\tau$ is defined from:
\begin{equation}
    {\rm d}\tau = -\varkappa_{\rm R}\rho {\rm d}z,
    \label{eq:tau_def}
\end{equation}
so that it increases from the surface to symmetry plane. Integrating equation~(\ref{eq:Pph}) gives:
\begin{equation}
    P_{\rm gas}(z_0) + \frac12 P_{\rm rad}(z_0) = \int_0^{2/3} \frac{\omega_{\rm K}^2 z_0}{\varkappa_{\rm R}(P_{\rm gas}, T(\tau))}\, {\rm d}\tau,
    \label{eq:P_bound_full}
\end{equation}
where we use the grey Eddington approximation for the temperature: $T(\tau) = T_{\rm eff} (1/2 + 3\tau/4)^{1/4}$. For a power-law opacity, the integral can be taken analytically~\citep{KetsarisShakura1998}, provided that coordinate $z$ hardly changes in the photosphere and equals to $z_0$.

The half-thickness of the disc $z_0$ is a free parameter. Thus, we have to set one additional boundary condition at the symmetry plane of the disc ($z=0$):
\begin{equation}
    Q(0) = 0, 
    \label{eq:Q_bound_another}
\end{equation}
which follows from the symmetry of the problem.

\subsection{Irradiation by central X-ray source}
\label{subsec:Irradiation-by-central-X-ray-source}

X-ray irradiation by the central accreting object (e.g., a neutron star) or by central parts of the accretion disc can be another source of heating, and even exceed the viscous heating at large radii. 

The spectrum of the incident radiation plays a major role. Soft X-rays are absorbed relatively high in the disc atmosphere and heat up the chromosphere-like layer, while photons with energy $> 3\,$keV can penetrate deep~\citep[][]{Suleimanov_etal1999}. If they are absorbed in the layers below the photosphere (where the optical depth for the disc own emission $\tau\sim2/3$), X-ray photons are thermalized and their energy is contributed to the flux outgoing from the photosphere. 

The surface temperature rises in presence of irradiation, and the new boundary condition can be written in the form
\begin{equation}
    T^4(z_0) = T_{\rm vis}^4 + T^4_{\rm irr}\, , 
    \label{eq:T_irr_bound}
\end{equation}
where irradiation temperature $T_{\rm irr}$ measures the additional heating by X-rays and will be defined differently in the two methods below. For irradiated discs we term \eqref{eq:T_bound} as `viscous temperature'
\begin{equation}
    T_{\rm vis} \equiv \left(\frac{Q_0}{\sigma_{\rm SB}}\right)^{1/4}.
\end{equation}

As we do not calculate an irradiated atmosphere model, we cannot calculate pressure from integral~(\ref{eq:P_bound_full}). Instead, following \citet{Tuchman_etal1990} and \citet{Hameury1998}, we assume that both the Rosseland opacity and $z$ are constant in the photosphere and take the value of pressure evaluated at $\tau=2/3$:
\begin{equation}
    P_{\rm gas}(z_0) + P_{\rm rad}(z_0) = \frac23\,\frac{\omega_{\rm K}^2 z_0}{\varkappa_{\rm R}(P_{\rm gas}(z_0), T(z_0))}.
    \label{eq:P_irr_bound}
\end{equation}

We use two ways to include irradiation. In the first method, only boundary conditions are changed~\citep[an approach similar to that by][]{Tuchman_etal1990, Dubus_etal1999}, while in the second one equations are altered as well.

\subsubsection{(i) First method}
In the first method, the boundary conditions on the temperature and pressure become (\ref{eq:T_irr_bound}) and (\ref{eq:P_irr_bound}). This roughly corresponds to all the heating caused by X-rays taking place at the photosphere level.

Irradiation temperature $T_{\rm irr}$ can be expressed in terms of irradiation parameter $C_{\rm irr}$:
\begin{equation}
    T^4_{\rm irr} = C_{\rm irr} \frac{L_{\rm X}}{4\pi\sigma_{\rm SB}r^2},
    \label{eq:Tirr_Cirr_relation}
\end{equation}
where $L_{\rm X}=\eta\dot{M}c^2$ is the X-ray luminosity of the central source.

\subsubsection{(ii) Second method}
\label{subsec:Irr_ii}
In the second method, X-ray radiation with arbitrary spectrum $F^\nu_X(\nu)$ penetrates into the disc and affects distributions of the energy flux $Q(z)$ and temperature $T(z)$.
In the disc the additional source of heating appears, so that Eq.~(\ref{eq:Q}) changes to the following form:
\begin{equation}
    \frac{{\rm d}Q}{{\rm d}z} = \frac{{\rm d}(Q_{\rm vis} + Q_{\rm irr})}{{\rm d}z} = \frac32\omega_{\rm K} \alpha P + \varepsilon,
    \label{eq:Q_irr}
\end{equation}
where $\varepsilon$ and $Q_{\rm irr}$ are the local heating rate of the disc through \hbox{X-ray} photons and the corresponding vertical energy flux. They are calculated from the analytical solution of radiation transfer equation for X-ray photons (see equations~(17)--(21) in \citet{Mescheryakov_etal2011} and Appendix~\ref{appendix:Irradiation-formulas}).
Accordingly, the boundary condition on the flux is changed: 
\begin{equation}
    Q(z_0) = Q_0 + Q_{\rm irr}(z_0).
    \label{eq:Q_irr_bound}
\end{equation}

The temperature and pressure boundary conditions are still  (\ref{eq:T_irr_bound}) and (\ref{eq:P_irr_bound}), respectively. 

While in the previous method the irradiation temperature or irradiation parameter can be an input parameter, now $T_{\rm irr}$ is calculated from calculated flux $Q_{\rm irr}$ \eqref{eq:Q_irr_app} at the photosphere level:
\begin{equation}
    \sigma_{\rm SB} T^4_{\rm irr}= Q_{\rm irr}(z_0).
    \label{eq:Tirr_Cirr_from_Qirr}
\end{equation}
 
The system has two free parameters: $z_0$ and the surface density of the disc $\Sigma_0$. Therefore, the code solves a two-parameter optimization problem and finds $(z_0, \Sigma_0)$. 
Contrary to the previous method, where the surface density of the disc is obtained on solving the equations, now one has to set the additional boundary condition, complementary to~(\ref{eq:Q_bound_another}):
\begin{equation}
    \Sigma(z=0) = \Sigma_0. \label{eq:Sigma_bound_another}
\end{equation}

We assume that X-ray radiation comes from the point-like central object, whose flux at distance $r$ is:
\begin{equation}
    F^\nu_X(\nu) = \frac{L_{\rm X}}{4\pi r^2}\, S(\nu),
    \label{eq:X-ray-flux}
\end{equation}
where $S(\nu)$ is the spectrum of incident X-ray flux (in units of $1/\rm Hz$ or $1/\rm keV$ and normalized to unity over a specified frequency range), and $L_{\rm X}$ is the X-ray luminosity of the central source. Both $S(\nu)$ and $L_{\rm X}$ can be set by user in the code (see Appendix~\ref{appendix:Brief-Code-description}).

\subsection{Calculation}
We have developed \texttt{Python~3} code that solves equations presented above. Code is open-source and available from GitHub\footnote{\url{https://github.com/AndreyTavleev/DiscVerSt}} (see its short description in Appendix~\ref{appendix:Brief-Code-description}). Several examples of calculated vertical structures can be found in Appendix~\ref{appendix:Vertical-structure} for different effective temperatures for cases with and without external irradiation.

We have checked the consistency of the code with results of some previous works. For analytic opacities the obtained vertical structure of un-irradiated disc agrees with results by \citet{KetsarisShakura1998} \citep[see][]{Tavleev_etal2019_inproc, Tavleev_etal2022_inproc}. The irradiated disc structure agrees with results by \citet{Mescheryakov_etal2011} and that by \citet{Dubus_etal1999, Tuchman_etal1990}, for corresponding methods.

\section{Radial structure}
\label{sec:Radial-structure}

\begin{figure*}
	\center{\includegraphics[width=0.8\linewidth]{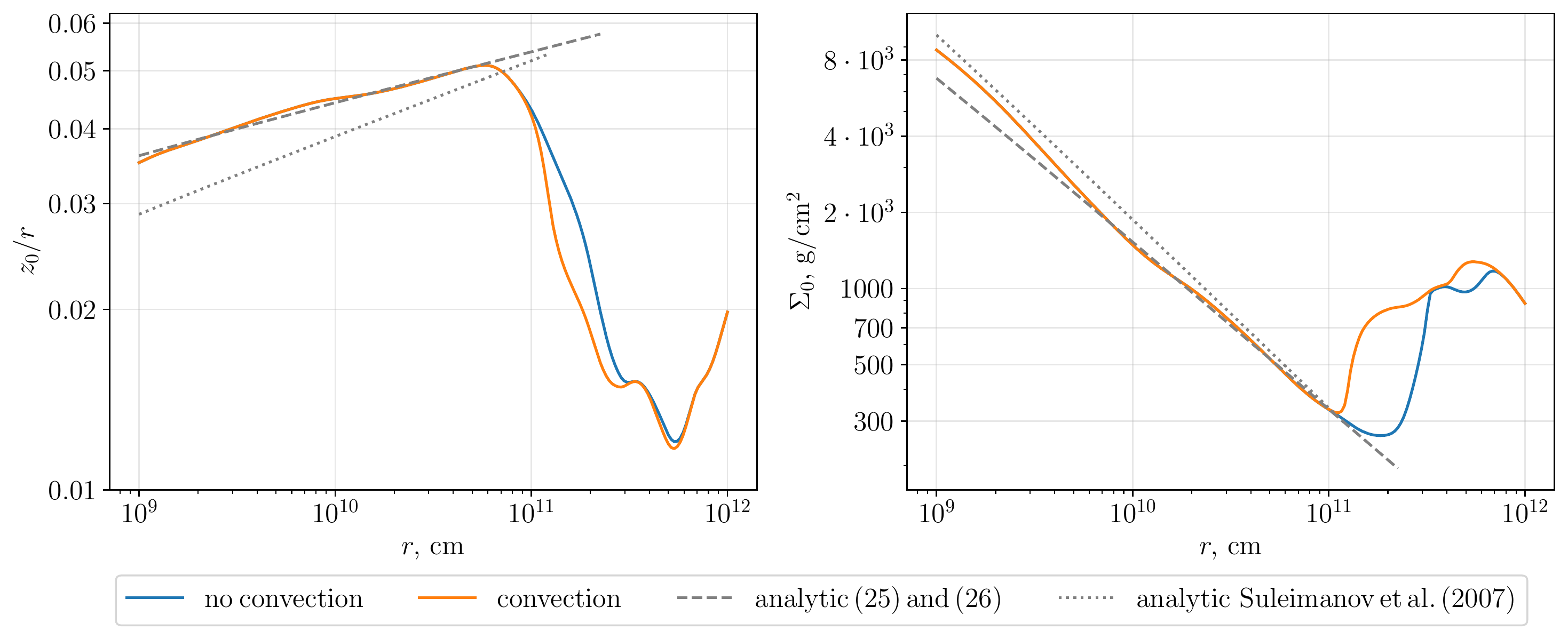}}
	\caption{Radial structure of disc with $M = 10\, M_{\sun}, \alpha = 0.1, \dot{M} = 10^{18} \rm g\, s^{-1}$, with and without convection together with theoretical approximations~(\ref{eq:Anal_formulas_1}--\ref{eq:Anal_formulas_2}) and analytical approximations by \citet{Suleimanov_etal2007}. Shown are the semi-thickness of disc $z_0/r$ and surface density $\Sigma_0$. It can be seen that convection thins the unstable part of the disc.}
	\label{fig:r_plots_conv_ZS}
\end{figure*}

Standard model of the disc accretion defines three radial zones~\citep{ShakuraSunyaev1973}. In zone A, the radiation pressure is greater than the gas pressure, and opacity is determined by scattering. In zone B, the gas pressure is greater than the radiation pressure, but opacity is still determined by scattering. Finally, in zone C, opacity is determined by the absorption processes, and gas pressure is much greater than the radiation pressure.

\citet{Suleimanov_etal2007} demonstrated that analytic radial dependencies of disc parameters in zones B and C could be written more accurately if solutions of vertical structure were taken into account. This approach relied on the method of vertical-structure calculation by \citet{KetsarisShakura1998} for analytical opacity coefficient and EoS, who introduced dimensionless $\Pi$-parameters. Here we write these parameters in a more general form:
\begin{equation}
\begin{split}
    \Pi_1 &\equiv \frac{\omega_{\rm K}^2\,z_0^2\,\rho_c}{P_c}, \qquad\qquad
    \Pi_2 \equiv \frac{\Sigma_0}{2z_0\,\rho_c}, \\
    \Pi_3 &\equiv \frac{3}{4}\frac{\alpha\,\omega_{\rm K}\,P_c\,\Sigma_0}{\rho_c Q_0}, \quad
    \Pi_4 \equiv \frac{3}{32}\left(\frac{T_{\rm eff}}{T_c}\right)^4\Sigma_0\,\varkappa_c.
\end{split} \label{eq:Pi}
\end{equation}

Here $P_c, T_c, \rho_c$ are the total pressure, temperature and bulk density in the symmetry plane. Values of $\Pi$-parameters can be found on solving the vertical structure for any opacity in a power-law form~\citep{KetsarisShakura1998,Malanchev_etal2017}. 

Knowing $\Pi$-values, one can obtain from (\ref{eq:Pi}) analytical formulas for the radial distribution of $z_0/r,\, \Sigma_0,\, T_c,\, \rho_c$. It was done for Kramers and Thomson opacity in \citet{Suleimanov_etal2007}, see also \citet{shakura_etal2018}. 

As we have already mentioned in relation with Figures~\ref{fig:Opacity_all} and \ref{fig:Opacity-diagram}, near the outer boundary of the hot disc (see the cyan dot on the olive curve in Fig.~\ref{fig:Opacity-diagram}), the opacity approximation $\varkappa_{\rm R}\sim\rho\,T^{-5/2}$ fits better the tabular opacity than the Kramers law ($\varkappa_{\rm R}\sim\rho\,T^{-7/2}$) does.
Let us substitute the EoS of ideal gas in (\ref{eq:Pi}) along with the opacity approximation formula obtained by \citet{BellLin1994} for hot disc regions, where opacity is determined by free-free and bound-free transitions. We obtain:
\begin{multline}
    z_0 / r = 0.0207\, m^{-13/36} \alpha^{-1/9}\, r_{10}^{1/12}\, \dot{M}_{17}^{1/6}\, f(r)^{1/6} \\ 
    \left(\frac\mu{0.6}\right)^{-13/36} \left(\frac{\varkappa_0}{\varkappa_0^*}\right)^{1/18} \Pi_z,
    \label{eq:Anal_formulas_1}
\end{multline}
\begin{multline}
    \Sigma_0 = 32\, m^{2/9}\, \alpha^{-7/9}\, r_{10}^{-2/3}\, \dot{M}_{17}^{2/3}\, f(r)^{2/3} \\
    \left(\frac\mu{0.6}\right)^{13/18} \left(\frac{\varkappa_0}{\varkappa_0^*}\right)^{-1/9} \Pi_{\Sigma} \,\, {\rm [g\, cm^{-2}]},
    \label{eq:Anal_formulas_2}
\end{multline}
\begin{multline}
    \rho_c = 7.8\cdot 10^{-8}\, m^{7/12}\, \alpha^{-2/3}\, r_{10}^{-7/4}\, \dot{M}_{17}^{1/2}\, f(r)^{1/2} \\ 
    \left(\frac\mu{0.6}\right)^{13/12} \left(\frac{\varkappa_0}{\varkappa_0^*}\right)^{-1/6} \Pi_{\rho} \,\, {\rm [g\, cm^{-3}]},
    \label{eq:Anal_formulas_3}
\end{multline}
\begin{multline}
    T_c = 4.1\cdot 10^4\, m^{5/18}\, \alpha^{-2/9}\, r_{10}^{-5/6}\, \dot{M}_{17}^{1/3}\, f(r)^{1/3} \\ 
    \left(\frac\mu{0.6}\right)^{5/18} \left(\frac{\varkappa_0}{\varkappa_0^*}\right)^{1/9} \Pi_{T} \,\, {\rm [K]}.
    \label{eq:Anal_formulas_4}
\end{multline}

Here:
\begin{equation}
\begin{split}
    &m \equiv \frac{M}{M_{\sun}}, \quad \dot{M}_{17} \equiv \frac{\dot{M}}{10^{17}\, {\rm g\, s^{-1}}}, \quad r_{10} \equiv \frac{r}{10^{10}\text{ cm}}, \\ 
    &\varkappa_0^* \equiv 1.5\cdot 10^{20} \text{ cm}^5\text{g}^{-2}\text{K}^{5/2}, \quad f(r)\equiv \frac{F}{\dot{M}h} = 1 - \sqrt{\frac{r_{\rm in}}{r}} + \frac{F_{\rm in}}{\dot{M}h}.
    \label{eq:Anal_formulas_5}
\end{split}
\end{equation}

Dimensionless parameters $\Pi_z, \Pi_{\Sigma}, \Pi_{\rho}, \Pi_{T}$ are almost constant in optically thick discs ($\tau\gtrsim10^4$) and are as follows:
\begin{equation}
\begin{split}
    \Pi_{z} &= \Pi_1^{17/36}\, \Pi_2^{-1/18}\, \Pi_3^{1/9}\, \Pi_4^{-1/18} \approx 2.6, \\
    \Pi_{\Sigma} &= \Pi_1^{1/18}\, \Pi_2^{1/9}\, \Pi_3^{7/9}\, \Pi_4^{1/9} \approx 1.049, \\
    \Pi_{\rho} &= \Pi_1^{-5/12}\, \Pi_2^{-5/6}\, \Pi_3^{2/3}\, \Pi_4^{1/6} \approx 0.771, \\
    \Pi_{T} &= \Pi_1^{-1/18}\, \Pi_2^{-1/9}\, \Pi_3^{2/9}\, \Pi_4^{-1/9} \approx 1.095.
\end{split}
\label{eq:Pi_combinations}
\end{equation}

\begin{figure*}
	\center{\includegraphics[width=1.0\linewidth]{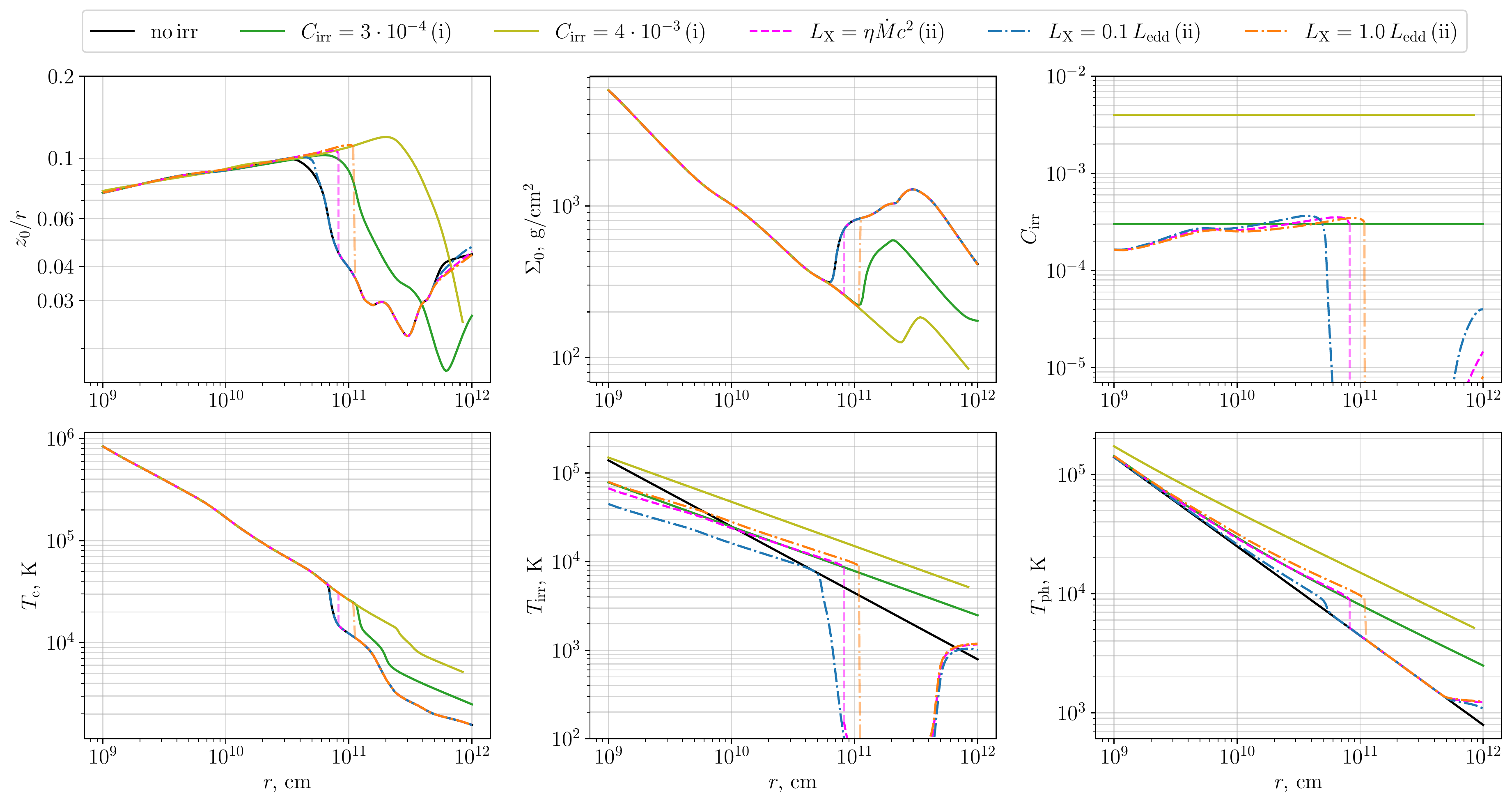}}
	\caption{Radial profiles of the vertical height $z_0/r$, surface density $\Sigma_0$, mid-plane temperature $T_c$, temperature at the photosphere $T_{\rm ph}$, irradiation temperature $T_{\rm irr}$ and irradiation parameter $C_{\rm irr}$ for un-irradiated disc and irradiated disc with irradiation scheme~(i) (see Sect.~\ref{subsec:Irradiation-by-central-X-ray-source}) for two $C_{\rm irr}$, and with advanced scheme~(ii) for different X-ray luminosities. Mass of central object $M = 1.4\, M_{\sun}, \alpha=0.1, \dot{M} = 10^{18}\,{\rm g\, s^{-1}}\approx 0.5 \,\dot{M}_{\rm edd}$. The black line in lower figures shows the viscous temperature $T_{\rm vis}$.}
	\label{fig:r_plots_irr}
\end{figure*}

\autoref{fig:r_plots_conv_ZS} presents the radial structure of solar disc with and without convection together with analytical approximations~(\ref{eq:Anal_formulas_1}--\ref{eq:Anal_formulas_2}) and $\Pi_z, \Pi_{\Sigma}$ values~(\ref{eq:Pi_combinations}), which are in good agreement with calculations in the stable region. One can see that Eq.~(\ref{eq:Anal_formulas_1}) better describes the $z_0/r$-profile in the outer part of the hot disc than the analytical approximation by~\citet{Suleimanov_etal2007} does. Moreover, this approximation is better even for hotter part of the disc. This is due to the influence of radiation pressure ($P_{\rm rad}/P_{\rm gas}\sim 0.2$ at $r=10^9\rm\, cm$), which causes the $z_0/r$-profile to ``shift'' higher compared to a calculation without taking into account radiation pressure. 

Overall, our analysis show that approximations~(\ref{eq:Anal_formulas_1})--(\ref{eq:Anal_formulas_4}) provide a reliable description (accuracy better than 15\%) for $z_0/r$- and $\Sigma_0$-profile in the outermost regions of the stationary hot disc if parameters fall in $M=1-10\, M_{\sun},\, \alpha=0.01-1.0,\, \dot{M}=10^{-3}-1.0 \,\dot{M}_{\rm edd}$. Specifically, these outermost regions are $(0.01-1) \times R_{\rm hot}$ for $z_0/r$ and $(0.1-1) \times R_{\rm hot}$ for $\Sigma_0$, where $R_{\rm hot}$ is the outer radius of the hot stable zone\footnote{$R_{\rm hot}$ can be found using \eqref{eq.TP_app} from $\dot M, \alpha$, and $M$.}. 
Furthermore, Eqs.~(\ref{eq:Anal_formulas_1})--(\ref{eq:Anal_formulas_4}) describe satisfactorily the quasi-stationary structure of outermost parts of an evolving fully-ionized disc, for example during an outburst, if function $f(r)$~from~(\ref{eq:Anal_formulas_5}) is properly adjusted~\citep[see figure~1.19 and table~1.5 in][]{shakura_etal2018}.

It should be kept in mind that there are intervals in the radial dependencies in Fig.~\ref{fig:r_plots_conv_ZS} that correspond to thermally-unstable solutions of the vertical structure. These intervals manifest themselves by sharp positive slope of the surface density radial profile. Thus, shown radial structure cannot hold longer than for a thermal time in unstable parts. Notice that account of the convection ``shifts'' the instability region to the smaller radii.

\subsection{Irradiated disc}

\autoref{fig:r_plots_irr} presents radial profiles of the relative semi-thickness $z_0/r$, surface density $\Sigma_0$, mid-plane temperature $T_c$, temperature at the photosphere $T_{\rm ph}$, irradiation temperature $T_{\rm irr}$ and irradiation parameter $C_{\rm irr}$ for un-irradiated and irradiated disc. Unless indicated otherwise, for the central X-ray luminosity $L_{\rm X}=\eta\dot{M}c^2$ we assume accretion efficiency $\eta=0.1$. Irradiation is taken into account by two methods (see Sect.~\ref{subsec:Irradiation-by-central-X-ray-source}: with scheme~(i) for two values of $C_{\rm irr}$ (the dark and light green lines) and with advanced scheme~(ii) for different X-ray luminosities (the blue, orange, and magenta lines).

In scheme~(ii), for spectrum of incident X-rays in expression~(\ref{eq:X-ray-flux}) we take:
\begin{equation}
    S(\nu)\propto \left(\frac{E}{kT_{\rm sp}}\right)^{-0.4} \exp\left(-\frac{E}{kT_{\rm sp}}\right)
    \label{eq:X-ray-spectrum}
\end{equation}
in the range $1-10\,\rm keV$, with $T_{\rm sp}=8\, \rm keV$~\citep{Mescheryakov_etal2011}. Note that spectrum $S(\nu)$ is the parameter of the code and can be set manually (see Appendix~\ref{appendix:Brief-Code-description}). The incident angle of external irradiation is assumed to be
\begin{equation}
    \cos\theta_0 \approx \frac{{\rm d} z_0}{{\rm d}r} - \frac{z_0}{r} = \frac{z_0}{r} \left(\frac{{\rm d}\ln z_0}{{\rm d}\ln r}-1\right) = \frac1{12} \frac{z_0}{r},
    \label{eq:cos_theta_irr}
\end{equation}
following the analytical approximation~(\ref{eq:Anal_formulas_1}) for $z_0/r$ in the case of a steady disc, when $f(r)\approx 1$.

It is known that $f(r)$ differs from $1$ for quasi-stationary hot zones of discs during outbursts~\citep[][figure~1.19]{shakura_etal2018} and its analytical form may be used in~\eqref{eq:cos_theta_irr}. On the other hand, the height where the X-rays are effectively intercepted may differ from $z_0$ remarkably. This can be due to scattering above the disc~\citep{Suleimanov_etal2007,Mescheryakov_etal2011}. Possible changes to (\ref{eq:cos_theta_irr}) are allowed in the code, see Appendix~\ref{appendix:Brief-Code-description}.

Interestingly, a physically reasonable result is reproduced even with the incident angle~(\ref{eq:cos_theta_irr}) for a shadowed zone beyond the hot ionized one. This happens because the calculated value of $C_{\rm irr}$ drops there, see also section~\ref{ssubsec:Cirr}.

\subsubsection{Comparing results of methods for irradiated disc}
\label{ssubsec:Irr-methods}

In confirmation with results of previous works~\citep[e.g.,][]{Dubus_etal1999}, strong irradiation keeps the disc in the hot state at farther distances, comparing to the case without irradiation.
We find that irradiation method slightly affects the hot disc size. For comparable values of $Q_{\rm irr}$, cf. the dark green and magenta lines in Fig.~\ref{fig:r_plots_irr}, the hot disc has very similar radial extension, which is seen in the panel for $\Sigma$ as a location where the surface density starts to rise going outwards. Additionally, there is difference in the relative thickness profiles: the disk calculated in scheme (i) becomes shadowed at smaller radius, whereas scheme (ii) yields shielding right at the hot zone radius.

Furthermore, stabilization of the disc's vertical structure by irradiation with $T_{\rm irr}\gtrsim 10^4\,\rm K$, previously found by~\citet{Tuchman_etal1990, Dubus_etal1999}, occurs in our calculations by either irradiation method. The temperature of the stability loss, which is actually $\lesssim 9000\,\rm K$, is investigated by us in detail in section~\ref{sec:Discussion}.

\subsubsection{Value of $C_{\rm irr}$}
\label{ssubsec:Cirr}

The irradiation parameter $C_{\rm irr}$ and irradiation temperature $T_{\rm irr}$ for advanced scheme~(ii), also shown in Fig.~\ref{fig:r_plots_irr}, can be calculated from the flux $Q_{\rm irr}$ (see~(\ref{eq:Tirr_Cirr_from_Qirr})~and~(\ref{eq:Qirr_Cirr_Tirr_app})). One can see that both $C_{\rm irr}$ and $T_{\rm irr}$ drop dramatically in the region where $T_{\rm irr} < 9000\,\rm K$. On comparing the curves for $C_{\rm irr}$ with disc profiles, we infer that the drop of $C_{\rm irr}$ is not due to a purely geometrical effect, since the cosine between incident rays and normal to the disc surface, which is set proportional to $z_0/r$ according to~(\ref{eq:cos_theta_irr}), decreases by only a few times. We deduce that the drop happens due to strong absorption of X-ray photons above the disc photosphere, so that the external irradiation hardly affects the disc vertical structure. This increased absorption is driven by a very high column density in the photosphere of the outer part of a disc (see the dashed line in Fig.~\ref{fig:C_irr_Sigma_ph}), which, in its turn, results from setting $\dot{M}=\rm const$ at all radii.

\begin{figure}
    \center{\includegraphics[width=1.0\linewidth]{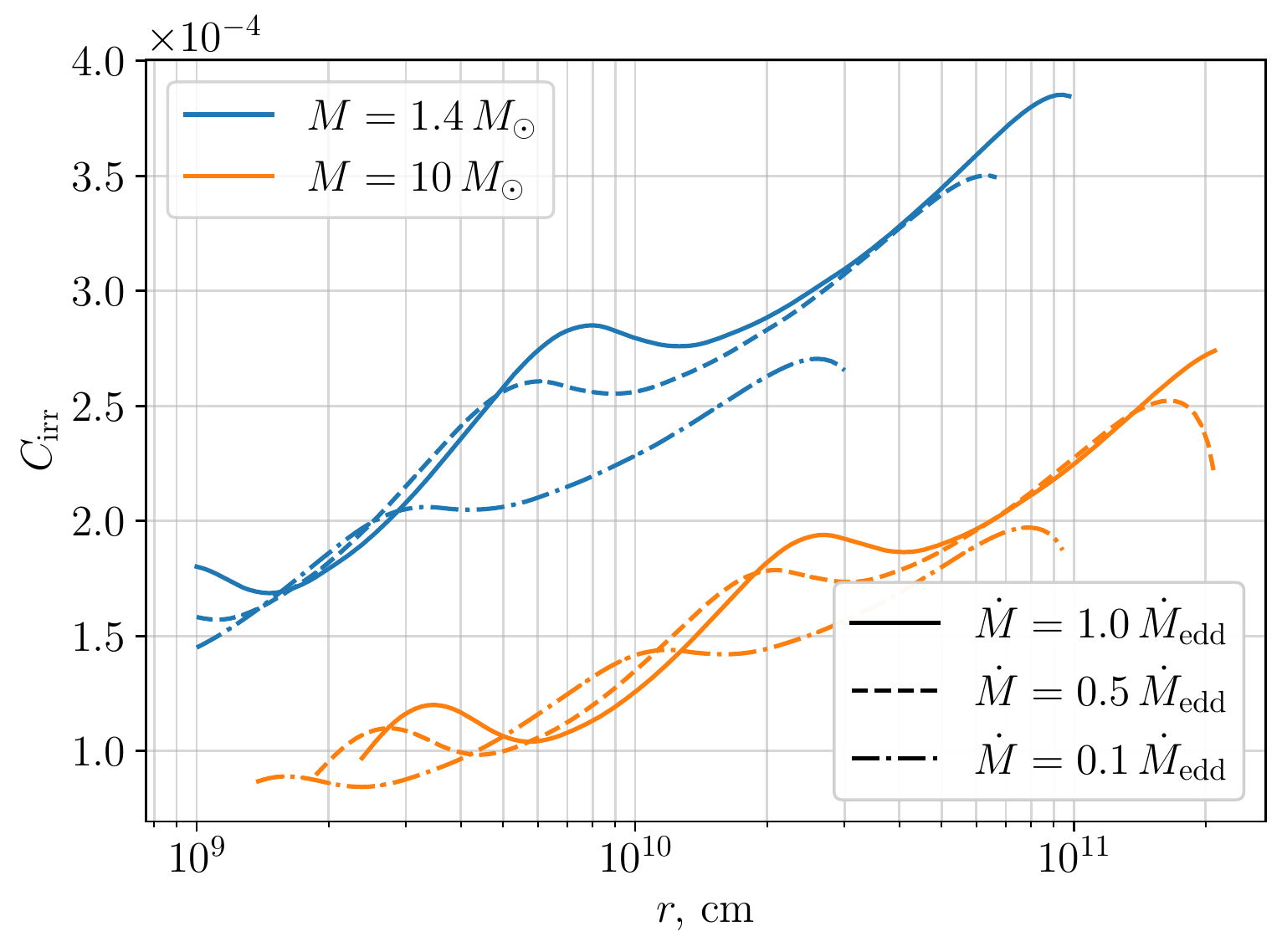}}
	\caption{Radial profile of irradiation parameter $C_{\rm irr}$ for two masses ($M = 1.4, 10\, M_{\sun}$) and three accretion rates ($\dot{M} = 0.1, 0.5, 1 \dot{M}_{\rm edd}$), $\alpha=0.1$.}
	\label{fig:C_irr}
\end{figure}

\begin{figure}
    \center{\includegraphics[width=1.0\linewidth]{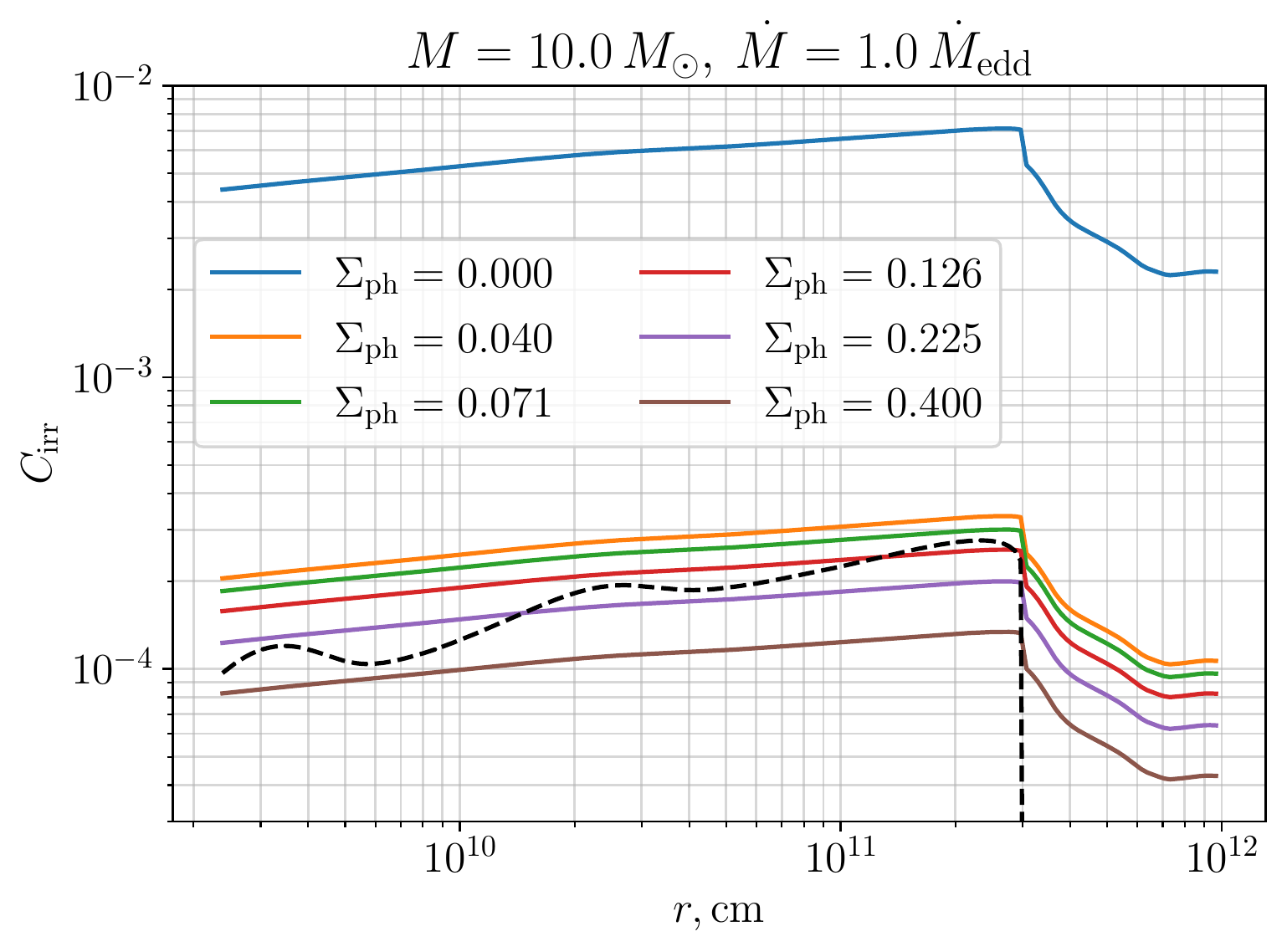}}
	\caption{Radial profile of $C_{\rm irr}$, when the column density of the photosphere layer above the disc surface $\Sigma_{\rm ph}=\rm const$ (coloured solid lines) in contrast with realistic $C_{\rm irr}$ profile (the black dashed line). System parameters are $M=10\, M_{\sun}, \dot{M}=1\,\dot{M}_{\rm edd}, \alpha=0.1$. The legends give the photosphere column density in g cm$^{-2}$. Drop of $C_{\rm irr}$ on the left occurs in the zone of the disc where the vertical structure is unstable.}
	\label{fig:C_irr_Sigma_ph}
\end{figure}

\begin{figure}
    \center{\includegraphics[width=1.0\linewidth]{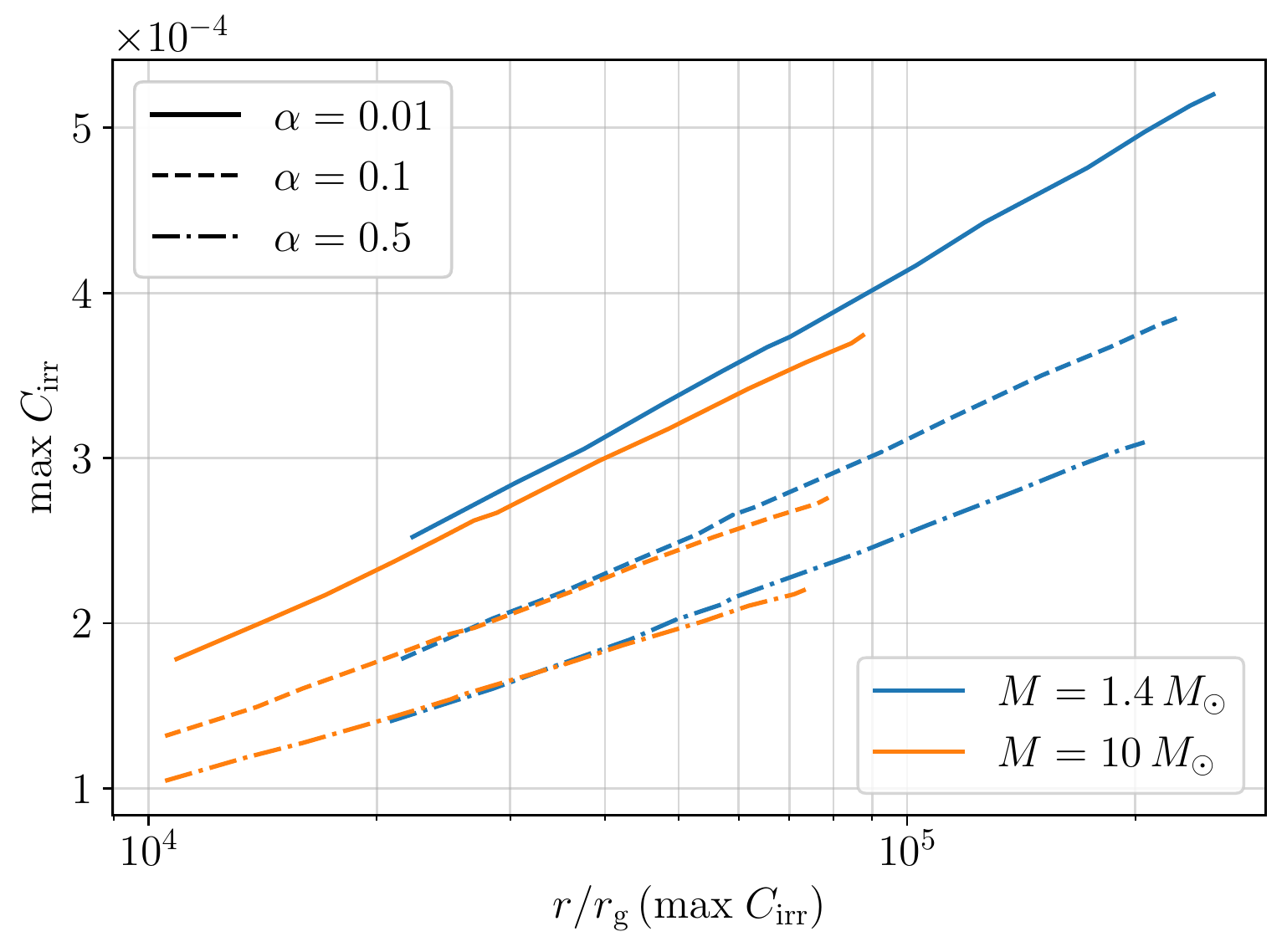}}
	\caption{Maximum $C_{\rm irr}$ for accretion rates $(10^{-2}-1) \dot{M}_{\rm edd}$, as function of the radius (in Schwarzschild radius) for two central source masses (different colours) and three $\alpha$-parameters (different styles). Irradiation is taken into account by method~(ii) and $L_{\rm X}=\eta\dot{M}c^2$.}
	\label{fig:max_C_irr}
\end{figure}

According to expression~(\ref{eq:Htot_disc}) and~(\ref{eq:Cirr_expression}), $C_{\rm irr}$ value depends on the X-ray spectrum, angle of incident rays and optical depth to X-rays above the level where the optical flux $\sigma_{\rm SB} T_{\rm vis}^4$ is formed (it is $\tau=2/3$ in our scheme). 
As we show in Appendix~\ref{appendix:Irradiation-formulas} for single-frequency X-ray photons, if the the column density of the photosphere layer above the disc surface $\Sigma_{\rm ph}=0$ and the total optical thickness of the disc $\tau_0 \gg 1$, one obtains analytically that $C_{\rm irr} = (1-A)\,\cos\theta_0$, where $A$ is the frequency-dependent albedo (see Eq.~(\ref{eq:Cirr_expression_simple}) and equation~A39 in \citet{Mescheryakov_etal2011}). 
This is in agreement with the previously proposed definition of $C_{\rm irr}$, according to which, for a point-like source, the irradiation parameter can be written as~\cite[e.g.,][]{Suleimanov_etal2007}
\begin{equation}
    C_{\rm irr} = (1-A) \frac{z_0}{r} \left(\frac{{\rm d} \ln z_0}{{\rm d} \ln r} - 1\right),
    \label{eq:C_irr_through_albedo}
\end{equation}
where $1-A$ is the fraction of incoming X-ray flux that is subject to thermalization. 

Fig.~\ref{fig:C_irr} shows radial dependencies of the irradiation parameter for different masses and accretion rates for the fixed spectrum (\ref{eq:X-ray-spectrum}). We stress that $C_{\rm irr}$ depends on the the upper boundary condition that defines $\Sigma_{\rm ph}$ and determines the pressure condition~(\ref{eq:P_irr_bound}). The dependence $\Sigma_{\rm ph}(r)$ is what drives the radial `wiggles' of $C_{\rm irr}$ in Fig.~\ref{fig:C_irr}: fixing photosphere column density to a constant value provides much more smooth $C_{\rm irr}$ behaviour, see Fig.~\ref{fig:C_irr_Sigma_ph}. An additional analysis showed that the dependence of the solid lines on radius in Fig.~\ref{fig:C_irr_Sigma_ph} came solely from the disc opening angle $z_0(r)/r$.

One could expect that $C_{\rm irr} \propto z_0/r$. If albedo $A$ in~(\ref{eq:C_irr_through_albedo}) is constant then a new parameter $\widetilde{C}_{\rm irr}$ can be suggested via
\begin{equation}
    C_{\rm irr} = \widetilde{C}_{\rm irr}\, \frac{z_0}{r},
\end{equation}
see, e.g., \citet{Lipunova_etal2022}. Fig.~\ref{fig:C_irr_z0r} in the Appendix shows the radial profile of $\widetilde{C}_{\rm irr}$. Indeed the variability of $\widetilde{C}_{\rm irr}$ with radius is less that that of ${C}_{\rm irr}$.
However, $\widetilde{C}_{\rm irr}$ appears to depend on the accretion rate, which, possibly, is due to a nontrivial dependence of $\Sigma_{\rm ph}$ and $A$ on $\dot{M}$. Auxiliary calculations have shown that, for varying value of the turbulent parameter~$\alpha$, from $0.01$ to $0.5$, the value of $\widetilde{C}_{\rm irr}$ changes by only about $\pm 5\%$.
However, the particular result is model-dependent and relies on the approximate boundary condition used in scheme (ii).

\section{S-curves}
\label{sec:S-curves}

\begin{figure*}
	\begin{minipage}{0.33\linewidth}
	\center{\includegraphics[width=1.0\linewidth]{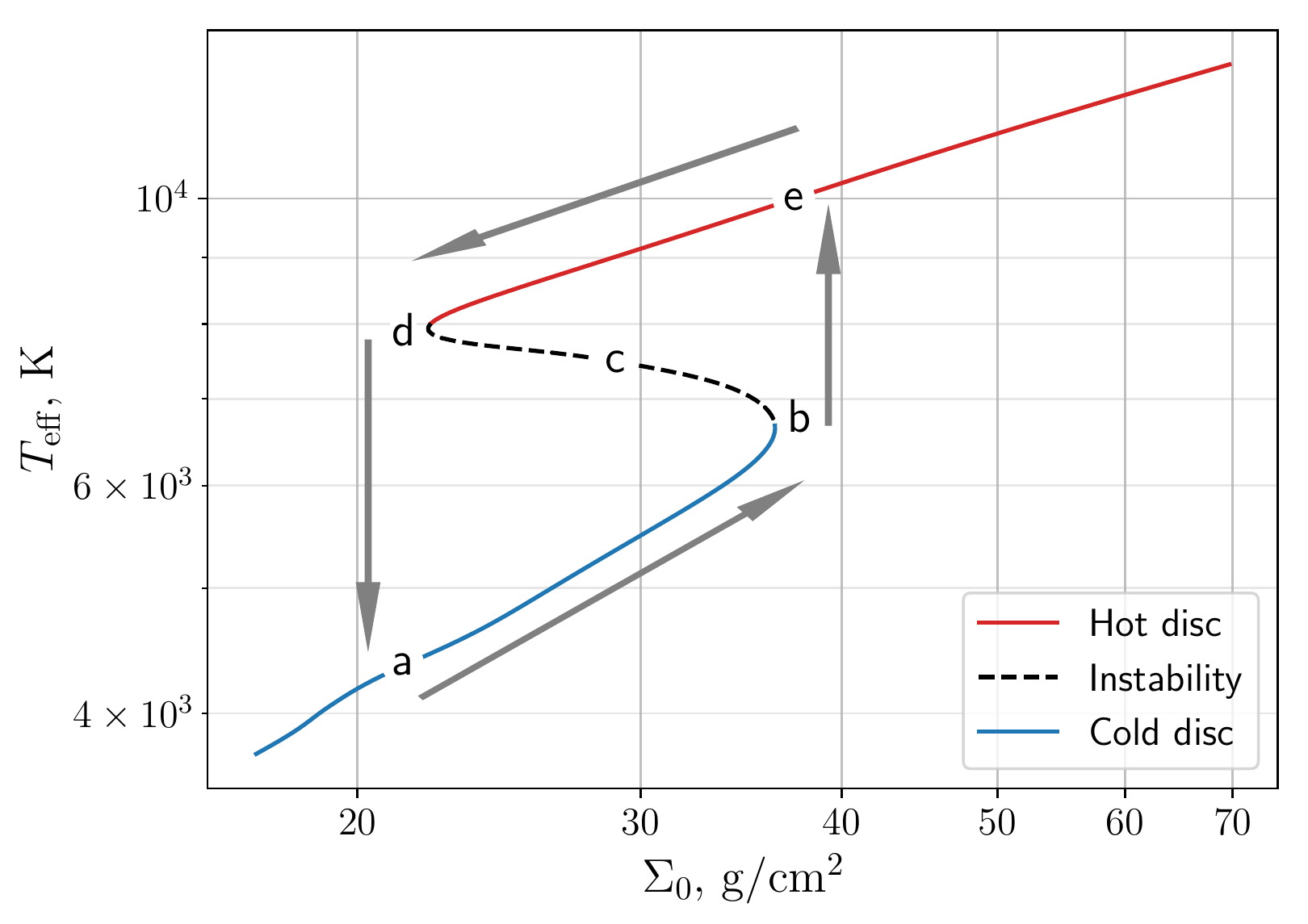}}
	\end{minipage}
	\begin{minipage}{0.33\linewidth}
	\center{\includegraphics[width=1.0\linewidth]{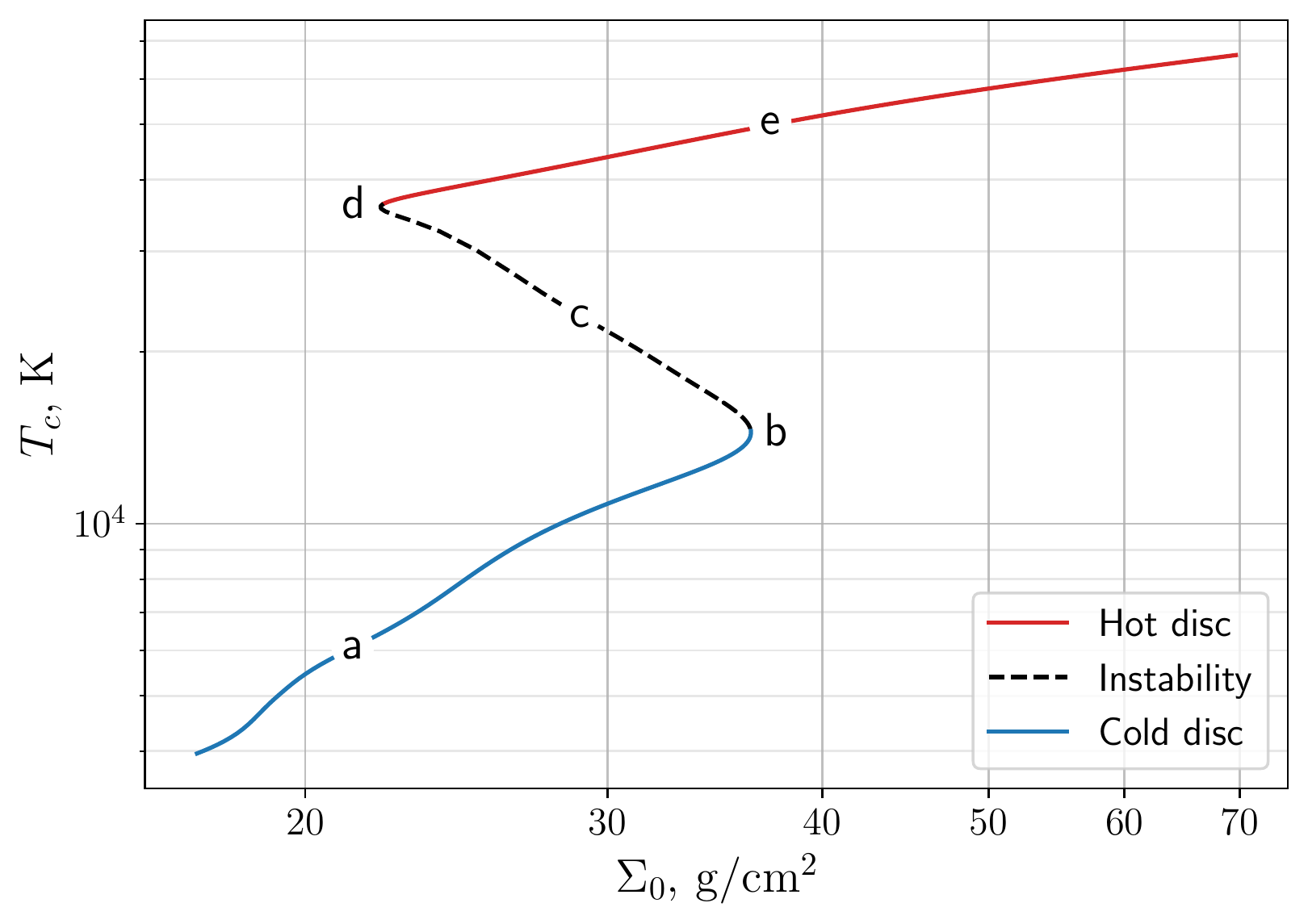}}
	\end{minipage}
	\begin{minipage}{0.33\linewidth}
	\center{\includegraphics[width=1.0\linewidth]{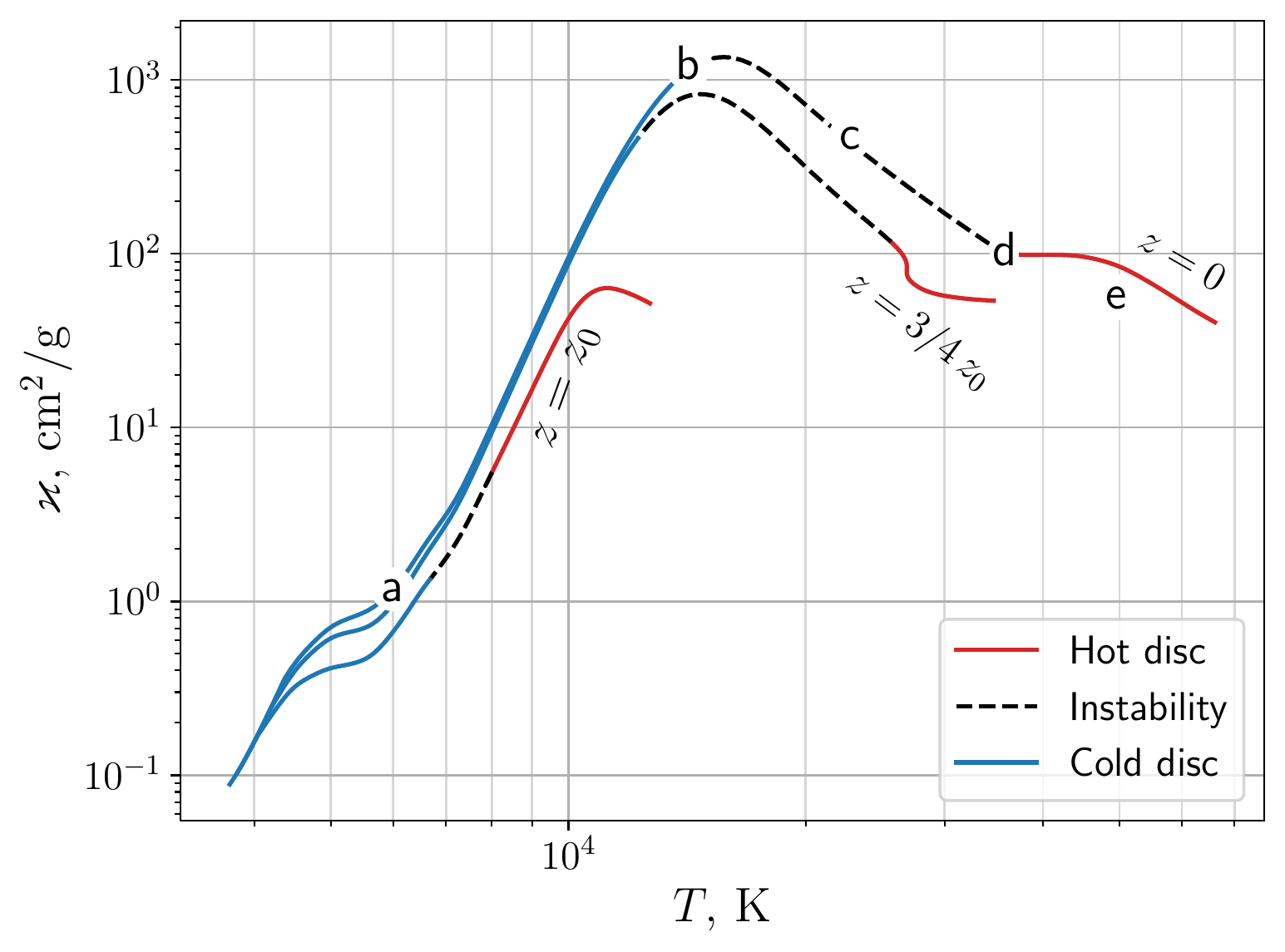}}
	\end{minipage}
    \caption{S-curve for $M = 10\, M_{\sun}, \alpha = 0.1,\, r = 10^{10}\, \rm cm$ and tabular opacities with solar chemical composition. The limit cycle is schematically shown by arrows. Also shown is the corresponding dependence of the symmetry plane temperature $T_c$ and the opacity coefficient as a function of temperature at the symmetry plane of the disc ($z=0$), at the disc surface ($z=z_0$) and in between ($z=3/4\,z_0$). The cold disc region, the region in which the instability takes place, and the region of the hot disc are marked with different style.}
\label{fig:S-curve}
\end{figure*}

\citet{Meyer1981} have established that dependencies $F - \Sigma_0$ (S-curves) show the disc instability: the branch of the S-curve with a negative slope represents solutions to the vertical structure equations which are viscously unstable, i.e. oscillations have to develop during characteristic time of order of the viscous time. \citet{Smak1982_small, Smak1984_part4} has showed that the viscously-unstable branch of the S-curve is also thermally unstable.
Note that since the quasi-stationary discs has an unambiguous relation between $F, \dot{M} \text{ and } T_{\rm eff}$ (see Eqs.~(\ref{eq:Q_bound}), (\ref{eq:T_bound}), (\ref{eq:Torque})), the S-curve can be drawn also in coordinates $\dot{M} - \Sigma_0$ and $T_{\rm eff} - \Sigma_0$.

An example of an S-curve calculated by our code is shown in Fig.~\ref{fig:S-curve}, the top left panel. Arrows shows schematically the direction of ring evolution when an outburst happens~\citep{FKR2002, Done_etal2007, KFM2008}: if in initially cold disc the surface density and temperature gradually rise (`a' $\rightarrow$ `b'), they eventually reach the critical point `b', where $T_{\rm eff} \sim 7000$~K and $T_{\rm c} \sim 13000$~K. The further temperature increase leads to a runaway heating: the thermal instability develops and brings the ring to the upper branch. 
There, a rapid viscous evolution occurs (`e' $\rightarrow$ `d'). The surface density decreases and the ring reaches the critical point `d' beyond which no stable `hot' solution is possible, and the ring transits to neutral `cold state'. Note a very different opacity dependence on the temperature in the upper and central layers of the disc for the unstable state (see the right panel of Fig.~\ref{fig:S-curve}, the dashed pieces; also \citet{Faulkner_etal1983_1}). This illustrates that irradiation of a hot decaying disc may inhibit an onset of instability by keeping the upper disc layer hot.

We have computed several thousand S-curves for $\alpha$ from $3\cdot10^{-4}$ to $0.7$, $M$ from $1\, M_{\sun}$ to $20\, M_{\sun}$ and $r$ from $7\cdot10^7$ cm to $5\cdot10^{11}$ cm and obtained the turning points $\Sigma^+$ and $\Sigma^-$, $T_{\rm eff}^+$ and $T_{\rm eff}^-$, $\dot{M}^+$ and $\dot{M}^-$, denoted below as `TP', which define the upper and lower S-curve turns, respectively. The chemical composition is assumed to be solar, and convective energy transport is taken into account. We fit the resulting TPs as:
\begin{equation}
    f(M, \alpha, r) = A\,\left(\frac{M}{M_{\sun}}\right)^{\beta}\, \alpha^{\gamma}\, \left(\frac{r}{10^{10}\,\text{cm}}\right)^{\delta},
    \label{eq.TP_app}
\end{equation}
where the parameters $A, \beta, \gamma, \delta$ and average relative uncertainty of TPs can be found in \autoref{tab:turn_points}. The average relative uncertainty~$\Delta$ is defined as
\begin{equation}
    \Delta = \left\langle\frac{|f(M,\alpha,r) - {\rm TP}|}{{\rm TP}}\right\rangle_{M,\alpha,r}.
\end{equation}

\begin{table}
\caption{Comparison of S-curve turning points (TP) with previous results. The minimum, maximum, and mean relative differences are shown. Negative difference value means that our TP approximation (\ref{eq.TP_app}) is less than the previous one.}
\centering
\begin{tabular}{cccc}
    TP & & (min;\,max) & mean \\
    \hline
    & L08$^a$ & $(15.2;\,33.8)\%$ & $24.5\%$ \\
    $\Sigma^+$ & CW84$^b$ & $(18.2;\,54.9)\%$ & $14.7\%$ \\
    & H98$^c$ & $(1.9;\,13.3)\%$ & $7.7\%$ \\
    \hline
    & L08 & $(-5.5;\,69.4)\%$ & $29.9\%$ \\
    $\Sigma^-$ & CW84 & $(-48.4;\,25.2)\%$ & $8.1\%$ \\
    & H98 & $(-17.1;\,7.8)\%$ & $4.9\%$ \\
    \hline
    $T_{\rm eff}^+$ & L08 & $(2.9;\,13.9)\%$ & $8.3\%$ \\
    \hline
    $T_{\rm eff}^-$ & L08 & $(6.4;\,21.4)\%$ & $13.1\%$ \\
    \hline
    \multicolumn{4}{l}{$^a$ \citet{Lasota_etal2008}} \\
    \multicolumn{4}{l}{$^b$ \citet{Cannizzo_Wheeler1984}} \\
    \multicolumn{4}{l}{$^c$ \citet{Hameury1998}} \\
    \end{tabular}
\label{tab:turn_points_diff}
\end{table}

\begin{table*}
\begin{tabular}{ccccccc}
     & $A$ & $\beta$ & $\gamma$ & $\delta$ & $\Delta$  \\
    \hline
    $\Sigma^+$ & $8.44 \pm 0.01 \,\rm g\, cm^{-2}$ & $-0.3674 \pm 0.0006$ & $-0.7821 \pm 0.0002$ & $1.1105 \pm 0.0002$ & $3.3\%$  \\
    $\Sigma^-$ & $11.87 \pm 0.03 \,\rm g\, cm^{-2}$ & $-0.3723 \pm 0.0009$ & $-0.8405 \pm 0.0003$ & $1.1223 \pm 0.0003$ & $5.4\%$  \\
    \hline
    $\dot{M}^+$ & $(1.027 \pm 0.003)\cdot 10^{16} \,\rm g\, s^{-1}$ & $-0.843 \pm 0.001$ & $-0.0193 \pm 0.0004$ & $2.6258 \pm 0.0003$ & $5.4\%$  \\
    $\dot{M}^-$ & $(5.065 \pm 0.016)\cdot 10^{15} \,\rm g\, s^{-1}$ & $-0.833 \pm 0.001$ & $0.0066 \pm 0.0004$ & $2.6038 \pm 0.0004$ & $6.5\%$  \\
    \hline
    $T_{\rm eff}^+$ & $7341 \pm 2 \,\rm K$ & $0.0290 \pm 0.0001$ & $-0.00484 \pm 0.00004$ & $-0.08426 \pm 0.00004$ & $0.7\%$  \\
    $T_{\rm eff}^-$ & $6152 \pm 4 \,\rm K$ & $0.0315 \pm 0.0002$ & $0.00165 \pm 0.00008$ & $-0.08977 \pm 0.00007$ & $1.3\%$  \\
    \hline
    \end{tabular}
    \caption{Values of the parameters (with standard deviations), which fit the S-curve turning points, where the '$+$' and '$-$' superscripts denote the upper and lower turning points, respectively. The right column contains the average relative uncertainty $\Delta$ of the turning points.}
\label{tab:turn_points}
\end{table*}

We find that these approximations of S-curves are generally close to the previous ones~\citep[][see \autoref{tab:turn_points_diff}]{Cannizzo_Wheeler1984, Hameury1998, Lasota_etal2008, Hameury2020_review}.
Taking into account that \citet{Liu_etal1997} have shown that the improved tabular opacities do not affect much the S-curves, we tend to conclude that uncertainties in $\Sigma_0$ are related to the slightly different boundary condition for pressure at the disc surface. The boundary condition~(\ref{eq:P_irr_bound}) is used in previous works, while we use~(\ref{eq:P_bound_full}) instead. Other possible source of uncertainty is the EoS tables. \citet{Lasota_etal2008} used tables from~\citep{Fontaine_etal1977}, while we use OPAL EoS tables~\citep{Rogers&Nayfonov2002}.

\begin{figure}
	\center{\includegraphics[width=1.0\linewidth]{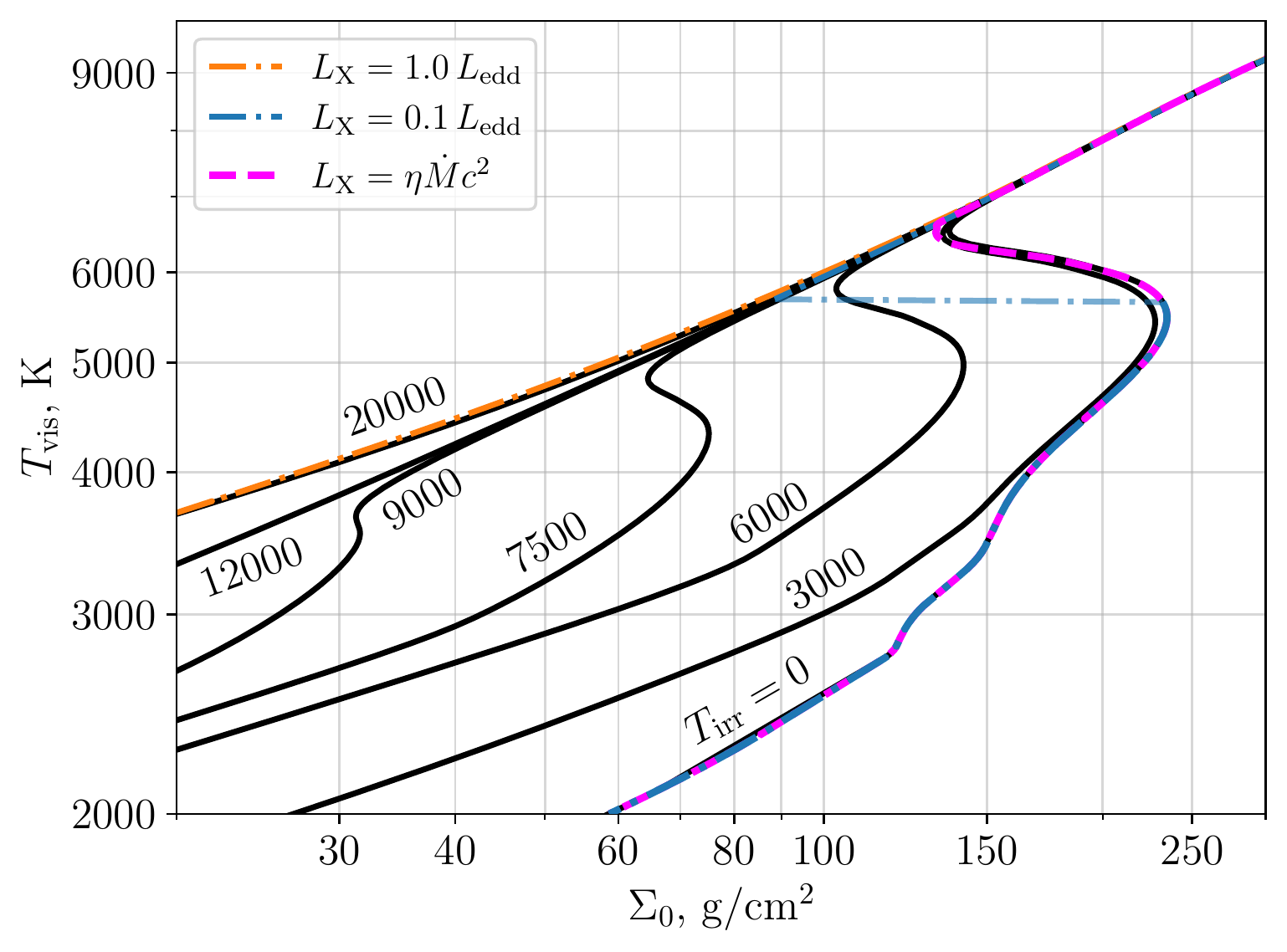}}
	\caption{S-curves for irradiated discs. Black curves are calculated through scheme~(i) with different irradiation temperatures $T_{\rm irr}$. Coloured lines are the curves with irradiation, calculated with advanced scheme~(ii). The magenta line corresponds to self-consistent luminosity of irradiation source $L_{\rm X} = \eta\dot{M}c^2$, while the blue and orange lines corresponds to luminosities $L_{\rm X} = 0.1, 1.0\, L_{\rm edd}$. All curves are calculated for $M = 1.4\, M_{\sun},\, r = 3\cdot10^{10} \, {\rm cm},\, \alpha = 0.1$ and solar chemical composition.}
	\label{fig:S-curve_irr}
\end{figure}

\subsection{Irradiation and S-curves}
\label{subsec:Irradiation-and-S-curves}

\autoref{fig:S-curve_irr} illustrates the influence of external irradiation on the disc stability. 
Considering the S-curves, calculated through scheme~(i) for different irradiation temperatures $T_{\rm irr}$, we infer that the unstable branch shrinks with the increase of $T_{\rm irr}$. For $T_{\rm irr}\gtrsim 10\,000\,\rm K$ the unstable branch disappears, so the strong irradiation stabilizes the disc, which is in agreement with the previous works~\citep[e.g.,][]{Tuchman_etal1990, Dubus_etal1999}.

In scheme (ii), the irradiation temperature $T_{\rm irr}$ varies along the upper branch of the S-curve for compliant luminosity (the magenta line). It hardly varies for the orange and blue line (with fixed $L_{\rm X}$), which is in accordance with $C_{\rm irr }$ being hardly dependent on $\dot M$, see Fig.~\ref{fig:C_irr}. On the lower and middle branch, $T_{\rm irr}$ in scheme (ii) is very low because significant part of X-rays are absorbed above the disc photosphere, see section \ref{ssubsec:Cirr}.

The disc calculated with advanced irradiation scheme~(ii) and self-consistent X-ray luminosity (magenta line) loses its stability when $T_{\rm irr} = T_{\rm irr,\, crit} \approx7500\,\rm K$, $T_{\rm eff} \approx 6600\,\rm K$ and $\dot{M} \approx 0.06\, \dot{M}_{\rm edd}$. This critical value $T_{\rm irr,\, crit}$ is lower than the one obtained in method~(i) and by~\citet{Tuchman_etal1990, Dubus_etal1999}. Furthermore, it depends on disc parameters, as we show in section~\ref{sec:Discussion}, see also Fig.~\ref{fig:T_irr_crit}. Note that in case with $L_{\rm X}=0.1\, L_{\rm edd}$ the negative slope branch is numerically unstable and cannot be calculated reliably, which leads to S-curve discontinuity.

\subsection{Influence of chemical composition, \texorpdfstring{$\bm\alpha$ parameter}, and convection on the shape of S-curves in X-ray transients}
\label{subsec:S-types}

\begin{figure}
	\center{\includegraphics[width=1.0\linewidth]{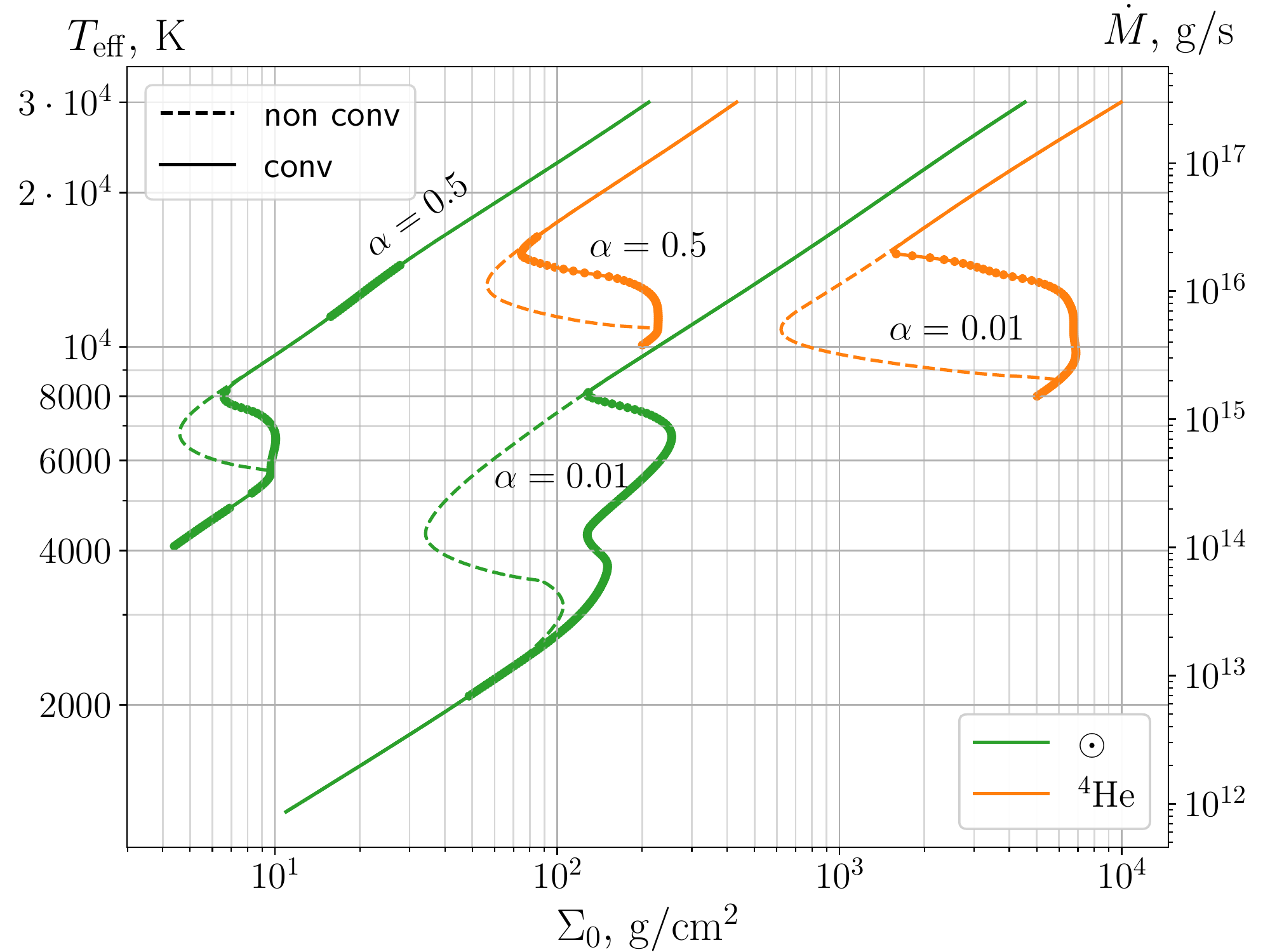}}
	\caption{S-curves, for different chemical composition, $\alpha$, with and without convection. All curves are calculated for $M = 10\, M_{\sun}$ and $r = 10^{10}\, \rm cm$. Bold point regions are the zones where the disc is convective (i.e. the condition for the existence of convection $\nabla_{\rm rad}>\nabla_{\rm ad}$ is fulfilled over an $\Sigma$-range of more than $50\%$). Note that only the optically thick branches of the curves are shown. For this reason, the curves for the helium disc do not show the area corresponding to the cold disc, since it is optically thin.}
	\label{fig:Experiment}
\end{figure}

\autoref{fig:Experiment} shows S-curves for different chemical composition and $\alpha$~parameter. Curves with and without convection are also shown. The dots mark the regions where convection in the disc dominates (i.e. the condition for the existence of convection $\nabla_{\rm rad}>\nabla_{\rm ad}$ is fulfilled in more than $50\%$ over $\Sigma$). It is seen that the disc is convective in the unstable region.

According to \citet{Faulkner_etal1983_1}, convection does not affect the very existence of instability. Indeed, we also obtain that, when the convection is ignored in the calculation, there is one unstable branch, regardless of the chemical composition or $\alpha$.

When convection is taken into account, the instability starts at higher accretion rates. For large $\alpha$, regions with convection are ``pulled'' towards large $\Sigma$, and an almost vertical interval at lower unstable branch is formed. For small $\alpha$, the convective branch splits into two unstable branches. In this case, the upper unstable branch is due to a peak in opacity related to the partial ionization of hydrogen. The lower branch is associated with convection~\citep{Cannizzo1992} and with the formation of molecular hydrogen~\citep{Smak1982_small}, see small peak in opacity (Fig.~\ref{fig:Opacity_all}) at $T\approx4000-5000 \, \rm K$. 
This `wiggle' is not usually associated with the outburst mechanism since the $\alpha$-value is believed to change only when the ionization degree is changing. At large~$\alpha$, the lower unstable branch does not appear, since the temperature does not reach such low values, at which the formation of molecules begins.

On the S-curve for helium disc, the instability is related with partial ionization of helium, so the corresponding temperatures are larger than for solar and hydrogen disc. At both large and small $\alpha$, the second unstable branch does not appear, only the main unstable branch is deformed.

These results for solar discs are consistent with the results in \citet{Cannizzo_etal1982}, where S-curves were investigated taking into account convection, which led to the appearance of additional kinks on the curve.

\section{Discussion}
\label{sec:Discussion}

Values of $C_{\rm irr}$, calculated by our code, are comparable or smaller than those suggested in previous works. For specific X-ray spectrum, calculated irradiation parameter $C_{\rm irr}$ is in the range $(1-5)\times 10^{-4}$ (Fig.~\ref{fig:max_C_irr}). 
\citet{Esin_etal2000} obtain $C_{\rm irr}\sim 0.004$ by analyzing the light curves of soft X-ray transient A0620-00 (1975), and $C_{\rm irr}\simeq0.0014$ for GRS~1124-68 (1991). Those values are consistent with estimate $C_{\rm irr}\sim (2-4)\times 10^{-3}$ found by~\citep{Jong_etal1996} for some Low-Mass X-ray Binaries. On the other hand, \citet{Suleimanov+2008} for the same two transients obtain $C_{\rm irr}\sim 7\times 10^{-4} $ and $C_{\rm irr}\sim 3\times 10^{-4} $, respectively.
\citet{Lipunova_etal2017} estimate that $C_{\rm irr}\sim (3-6)\times 10^{-4}$ using optical data of \hbox{4U~1543$-$47}~(outburst of 2002). Assuming the DIM model, \citet{Coriat+2012} have analysed transient and persistent X-ray sources with neutron stars and black holes and concluded that irradiation parameter lies in the interval $10^{-3}-10^{-2}$. This is an order of magnitude higher than the values obtained by us.

There is a physical reason of why our irradiation scheme provides the lower limit on $C_{\rm irr}$. The present scheme involves neither additional heating of the disc by soft-X-rays-heated upper layers of the photosphere (above $\tau=2/3$) nor the increased X-ray flux due to scattering in even higher and hotter corona \citep[see][]{Suleimanov_etal2007, Mescheryakov_etal2011}.
For supersoft X-ray sources, \citet{Suleimanov_etal2003} suggest that relatively dense blobs immersed in a corona enable multiple X-ray or far-UV scattering which leads to observed large optical and UV fluxes. The same mechanism might increase the irradiation parameter $C_{\rm irr}$.

The self-irradiation plays a crucial role in an outburst dynamics, since the
hot disc size affects directly the duration of an outburst~\citep[see e.g.,][]{King-Ritter1998}. 
It was shown before that there is a minimum irradiation temperature that ensures the disc stability~\citep{Tuchman_etal1990, Dubus_etal1999}, estimated as $(9-10)\times 10^3\,\rm K$. While the actual $C_{\rm irr}$ can be higher comparing to the values we find, the critical disc irradiation temperature, which switches on/off ionization instability, can be reliably obtained, since it depends not on $C_{\rm irr}$ but on how much the irradiation flux exceeds the internal viscous one.

\begin{figure}
    \center{\includegraphics[width=1.0\linewidth]{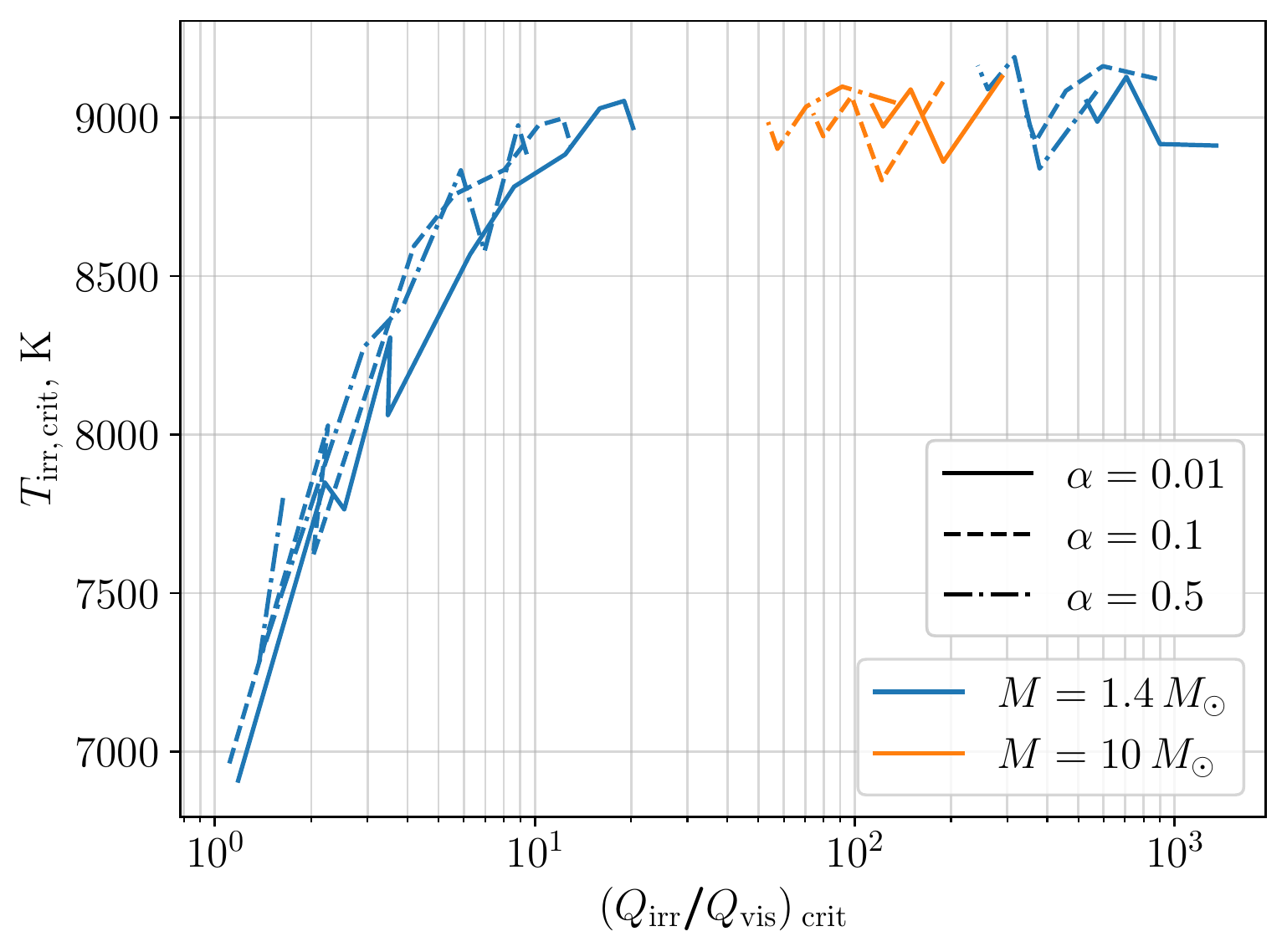}}
	\caption{Critical value of irradiation temperature $T_{\rm irr,\, crit}$, above which the disc is stable, as function of the ratio between irradiation and viscous heat $Q_{\rm irr} / Q_{\rm vis}$ for two central source masses (different colours) and three $\alpha$-parameters (different styles). Irradiation is taken into account by method~(ii). Note that weak irradiation does not affect the disc around $10\, M_{\sun}$, so there is no $T_{\rm irr,\, crit}$ in this case.}
	\label{fig:T_irr_crit}
\end{figure}

In Fig.~\ref{fig:T_irr_crit} we show dependence of the critical irradiation temperature on the ratio of the irradiation to viscous heat. For this we have calculated numerous models of irradiated disc with scheme~(ii), with arbitrary values of the central flux (to cover the scenarios of enhanced values of $C_{\rm irr}$).

For strongly illuminated discs, when $Q_{\rm irr} > Q_{\rm vis}$, irradiation controls the disc size. This occurs for big discs and sufficiently high $C_{\rm irr}$, suitably illustrated by a formula from \citet{Suleimanov_etal2007}:
\begin{equation}
\frac{Q_{\rm irr}}{Q_{\rm vis}} = \frac 43 \, \eta \, C_{\rm irr} \, \frac{r}{r_{\rm g}}.
\label{eq:qratio}
\end{equation}
Thus, at radii $r > 3/4\, r_{\rm g} / (\eta \, C_{\rm irr})$ the stability condition is $T_{\rm irr} >T_{\rm irr, crit} $, see Fig.~\ref{fig:T_irr_crit}.

In the opposite case, if $Q_{\rm irr} < Q_{\rm vis}$, irradiation does not affect the disc structure and the disc stability: unstable state is triggered at the radius where the effective temperature lowers to $T_{\rm eff}^{+}$. This can happen in the case of relatively small discs or small $C_{\rm irr}$.

\subsection{Code limitations}

The code has been tested for discs with $T_{\rm eff}\sim (10^3-10^6) \rm\, K$ around stellar-mass central sources. The code can also be used to calculate discs around supermassive black holes (SMBH), where a specific zone $B^*$ appears, a mixture of $A$ and $C$ disc zones~\citep[][see also Sect.~\ref{sec:Radial-structure}]{Burderi_etal1998}. Note that the disc self-gravity becomes important when
\begin{equation}
    \frac{2\pi G\Sigma}{\omega_{\rm K}^2 z_0} > 1\,.
    \label{eq:self_gravity}
\end{equation}
Inequality~(\ref{eq:self_gravity}) is not checked automatically in the code, so this needs to be checked separately. 

When $P_{\rm rad}\gtrsim P_{\rm gas}$, a thermal-viscous instability~\citep{Lightman_etal1974, ShakuraSunyaev1976} appears and, accordingly, the solution of system \eqref{eq:P}--\eqref{eq:Sigma} becomes problematic. Additionally, at high accretion rates, the local energy balance breaks: the energy is effectively transported in radial direction with the advected portion being of order of $(z_0/r)^2$. Our code does not calculate such discs.

While we present radial profiles and S-curves with a constant $\alpha$-parameter, it should be noted that $\alpha$ is expected to vary. For example, \citet{Smak1984_part4} showed that $\alpha$ should be lower by a factor of several in the `cold' zone where $T_{\rm eff}<5000\,\rm K$. The change in the turbulent parameter is dictated by the need to produce outbursts of observed amplitude~\citep[see e.g.,][]{Hameury_etal2009}. Observations show that $\alpha\sim0.1-1$ for the hot disc and $\alpha\sim0.01$ for the cold disc~\citep[see, e.g.,][]{Cannizzo_etal1988, suleimanov_etal2008, Kotko_etal2012,VISCOS2019, Tetarenko18}. 

MRI simulations provide intriguing details~\citep[see, e.g,][]{Hirose_etal2014, Scepi_etal2018, Jiang_etal2020}: $\alpha$ is higher near the upper turn of S-curve. The mechanism for such an increase is likely convection associated with the hydrogen ionization transition. In regions with high temperature, where convection does not occur, $\alpha$ remains low. However, there is no simple relationship between $\alpha$ and the strength of convection~\citep{Scepi_etal2018}.

Equation~(\ref{eq:Q}) of the viscous heating implies the local version of the $\alpha$-prescription for viscosity, where the \textit{local} tensor of viscous stress $w_{r\varphi}$ is proportional to the total pressure. However, this is not always a good approximation, particularly in the layers where $\tau<1$~\citep{Shaviv_etal1986}. An alternative form of $\alpha$-viscosity is the global form, where the \textit{vertically integrated} viscous stress $W_{r\varphi}$ is proportional to the vertically averaged total pressure. This form is used, e.g., in calculations of disc spectra~\citep{Hubeny_etal2021}.

Another important point is the pressure boundary condition, which is very approximate due to a simple atmospheric model used. Its proper determination requires the accurate calculation of the disc spectrum~\citep[see, e.g.,][]{Hubeny_etal2021}. We treat the atmosphere in the Eddington approximation and use the boundary condition for pressure (\ref{eq:P_bound_full}) or (\ref{eq:P_irr_bound}). These two approaches give boundary pressure $70\%$ apart from each other in the unstable convective region, leading to differences in values at the turning points~(see Sect.~\ref{sec:S-curves}).

\section{Summary}
\label{sec:Summary}

Calculation of the vertical structure of accretion discs is necessary to understand stability properties of accretion discs and to reconstruct the light curves of X-ray transients. 
Large range of physical conditions over a disc or various chemical composition in different objects require a numerical approach of calculating disc parameters, which is fast and flexible at the same time. Our first open numerical code for the model of the vertical structure with different types of equation of state and opacity laws, including tabular values, takes into account both radiative and convective energy transport and external X-ray irradiation. 

Using the results of numerical models, we obtain analytical formulas for radial dependencies of disc parameters using \citet{BellLin1994} power-law approximation of opacity coefficient in the high-temperature plasma. These formulas can be applied in the outermost parts of a hot ionized accretion disc around a stellar-mass compact object.

We analyse stability criteria for our disc model. For this, analytical approximations for the S-curve turning points are obtained. Comparing to previous results, variations in $\Sigma_0$ turning points are explained by slightly different boundary condition for pressure at the disc surface and different EoS tables.

For a case with external X-ray irradiation, stabilization of the disc’s vertical structure at $T_{\rm irr} > 10^4\,\rm K$, previously found by \citet{Tuchman_etal1990, Dubus_etal1999}, occurs in our models as well. Using the advanced scheme of calculation of irradiation disc vertical structure, we refine the critical value of $T_{\rm irr}$ and find that it changes in range $6900-9000\,\rm K$. We propose its unique dependence on the ratio between irradiation and viscous heat (Fig.\ref{fig:T_irr_crit}).
In addition, we calculate values of self-irradiation parameter $C_{\rm irr}$ in our model. They represent lower estimates of the actual $C_{\rm irr}$ in X-ray transients, since additional heating and/or scattering from the hot layers above the disc photosphere should enhance the heating effect.

\section*{Acknowledgements}

The authors are grateful to Valery Suleimanov for discussion and the referee for helpful suggestions. The work was supported by the RSF grant 21-12-00141.

\section*{Data Availability}
\label{sec:Data-Availability}

Code for calculating the vertical structure is open-source and available from GitHub\footnote{\url{https://github.com/AndreyTavleev/DiscVerSt}}.

The pre-calculated S-curve turning points ($\Sigma^+$ and $\Sigma^-$, $T_{\rm eff}^+$ and $T_{\rm eff}^-$, $\dot{M}^+$ and $\dot{M}^-$) are available~\footnote{\url{https://doi.org/10.5281/zenodo.7361425}} for $20$~linearly scaled values of $M$ from $1\, M_{\sun}$ to $20\, M_{\sun}$, 20~logarithmically scaled values of $\alpha$ from $3\cdot10^{-4}$ to $0.7$, 20~logarithmically scaled values of $r$ from $7\cdot10^7$ cm to $5\cdot10^{11}$ cm.




\bibliographystyle{mnras}
\bibliography{references}



\appendix

\section{Irradiation formulas}
\label{appendix:Irradiation-formulas}

In this section we write the formulas that describe the irradiation terms in the advanced irradiation scheme~(ii), see Sect.~\ref{subsec:Irr_ii}.

Assume that the disc is irradiated by external X-rays with spectral flux $F_X^\nu(\nu)$. The angle between the direction of incidence of the X-ray photons and the inward normal to the disc layer surface is $\theta_0$, while the cosine of this angle we denote as $\zeta_0=\cos\theta_0$.

The X-ray photons are scattered, absorbed and thermalized in the disc and can serve as additional heating source. The scattering in the medium is assumed to be coherent (Thomson scattering, $\sigma=\sigma_{\rm T}$), and the opacity coefficient $\varkappa^\nu$ for X-rays is determined by photoabsorption for a cold gas~\citep{MorrisonMcCammon1983}.

The mean intensity $J^\nu_{\rm tot}$ and flux $H^\nu_{\rm tot}$ of both primary and scattered X-ray photons in the disc at some depth with corresponding $\tau_\nu$ at frequency $\nu$ can be found by solving the transfer equation in plane-parallel approximation~\citep{Mescheryakov_etal2011}:
\begin{multline}
    J^\nu_{\rm tot}(\tau^\nu, \nu) = \frac{F^\nu_X}{4\pi}\bigg\{C^\nu\left[e^{-k\tau^\nu} + e^{-k(\tau^\nu_0-\tau^\nu)}\right] + \\
    +(1-D^\nu)\left[ e^{-\tau^\nu/\zeta_0} + e^{-(\tau^\nu_0 - \tau^\nu)/\zeta_0} \right] \bigg\}, 
    \label{eq:Jtot_disc}
\end{multline}
\begin{multline}
    H^\nu_{\rm tot}(\tau^\nu, \nu) = F^\nu_X\bigg\{\frac{k\,C^\nu}{3}\left[e^{-k\tau^\nu} - e^{-k(\tau^\nu_0-\tau^\nu)}\right] + \\ 
    +\left(\zeta_0-\frac{D^\nu}{3\,\zeta_0}\right)\left[e^{-\tau^\nu/\zeta_0} - e^{-(\tau^\nu_0 - \tau^\nu)/\zeta_0} \right]\bigg\},
    \label{eq:Htot_disc}
\end{multline}
where $\tau^\nu_0$ is the total optical depth of the disc in the vertical direction for X-ray radiation at frequency $\nu$, $\tau^\nu=\Sigma(\sigma+\varkappa^\nu)/2$, $\varkappa^\nu$ is the absorption coefficient for X-ray photons, $\sigma$ is the scattering coefficient, $k=\sqrt{3(1-\lambda)}$ and $\lambda={\sigma}/({\sigma+\varkappa^\nu})$ is the single-scattering albedo. Formulas for $C^\nu$ and $D^\nu$ can be found in \citet{Mescheryakov_etal2011}.

The additional heating of the disc by X-ray photons of a given frequency $\varepsilon^\nu$ is proportional to their mean intensity:
\begin{equation}
\varepsilon^\nu = 4\pi\rho\varkappa^\nu J^\nu_{\rm tot}.
\end{equation}

The local energy release in the disc through its irradiation by X-ray photons is
\begin{equation}
\varepsilon = \int_0^\infty \varepsilon^\nu \,{\rm d}\nu = 
4\pi\rho \int_0^\infty \varkappa^\nu J^\nu_{\rm tot}\, {\rm d}\nu.
\end{equation} 

The flux $H^\nu_{\rm tot}$ is calculated for all solid angles and it takes into account photons coming into the disc from outside minus those escaping the disc without absorption. Thus, the total heating of the disc from the disc surface to the central plane through its irradiation is
\begin{equation}
Q_{\rm irr}(z_0) = \int_0^\infty H^\nu_{\rm tot}(\tau^\nu_{\rm ph}, \nu)\, {\rm d}\nu,
\label{eq:Q_irr_app}
\end{equation}
where $\tau^\nu_{\rm ph} = (\sigma+\varkappa^\nu)\Sigma_{\rm ph}$ is the optical depth of the photosphere layers above the disc surface, $\Sigma_{\rm ph}$ is the corresponding column density. To find it we can write (cf.~(\ref{eq:Pph}-\ref{eq:tau_def}))
\begin{equation}
{\rm d}\Sigma_{\rm ph} = -\rho{\rm d}z = \frac{{\rm d}\tau}{\varkappa_{\rm R}}
\end{equation}
and take the value, evaluated at $z_0$, which corresponds to $\tau=2/3$:
\begin{equation}
    \Sigma_{\rm ph} = \frac23\,\frac1{\varkappa_{\rm R}(P_{\rm gas}(z_0), T(z_0))} = \frac{P_{\rm gas}(z_0) + P_{\rm rad}(z_0)}{\omega_{\rm K}^2 z_0}.
\end{equation}

It should be noted that photospheric column density $\Sigma_{\rm ph}$ is not included into the surface density $\Sigma_0$ of the disc when we calculate S-curves.

The irradiation temperature and irradiation parameter can be found from the irradiation flux \eqref{eq:Q_irr_app}:
\begin{equation}
    Q_{\rm irr}(z_0) = \sigma_{\rm SB} T^4_{\rm irr} = C_{\rm irr} \frac{L_{\rm X}}{4\pi r^2},
    \label{eq:Qirr_Cirr_Tirr_app}
\end{equation}
where $L_{\rm X}$ is the X-ray luminosity of the central source.

Notice that $\varepsilon$ is the function of $\Sigma$, that is, the function of the vertical coordinate $z$. The total X-ray optical depth is $\tau_0^\nu=(\sigma+\varkappa^\nu)(\Sigma_0+2\cdot\Sigma_{\rm ph})$. Therefore, irradiation terms $\varepsilon$ and $Q_{\rm irr}$ (as well as $T_{\rm irr}$ and $C_{\rm irr}$) contain the surface density $\Sigma_0$ as an additional free parameter, so the system of equations for the disc vertical structure in irradiation scheme~(ii) have two free parameters: $z_0$ and $\Sigma_0$.

Using (\ref{eq:Htot_disc}), (\ref{eq:Q_irr_app}), (\ref{eq:Qirr_Cirr_Tirr_app}) and~(\ref{eq:X-ray-flux}), we can obtain exact formula:
\begin{equation}
    C_{\rm irr} = \frac{\int_0^{\infty} F^\nu_X\left\{...\right\}\, {\rm d}\nu}{\int_0^{\infty} F^\nu_X \, {\rm d}\nu} = \int_0^{\infty} S(\nu) \left\{...\right\} \,{\rm d}\nu,
    \label{eq:Cirr_expression}
\end{equation}
where expression in $\left\{...\right\}$ is the one from the~(\ref{eq:Htot_disc}). For a very optically thick disc with $\tau^\nu_0 \gg 1$, exponential terms with $\tau^\nu_0$ tend to zero, and it can be shown that $C_{\rm irr}\propto\zeta_0$. If additionally we assume $\Sigma_{\rm ph} = 0$, then $\tau^\nu_{\rm ph}=0$, and we obtain
\begin{equation}
    C_{\rm irr} = \left( 1 - \int_0^\infty S(\nu) \frac{3\lambda}{(1+k\zeta_0)(3+2k)}\,{\rm d}\nu \right)\zeta_0,
    \label{eq:Cirr_expression_simple}
\end{equation}
or, for single-frequency incoming X-ray photons, $C_{\rm irr} = (1-A)\zeta_0$, where the frequency-dependent albedo $A$ is defined following \citet{Mescheryakov_etal2011}.

Moreover, one could introduce a notion of a spectrum-integrated albedo:
\begin{equation}
A^* = \int_0^\infty S(\nu) \frac{3\lambda}{(1+k\zeta_0)(3+2k)}\,{\rm d}\nu.
\end{equation}
\begin{figure}
    \center{\includegraphics[width=1.0\linewidth]{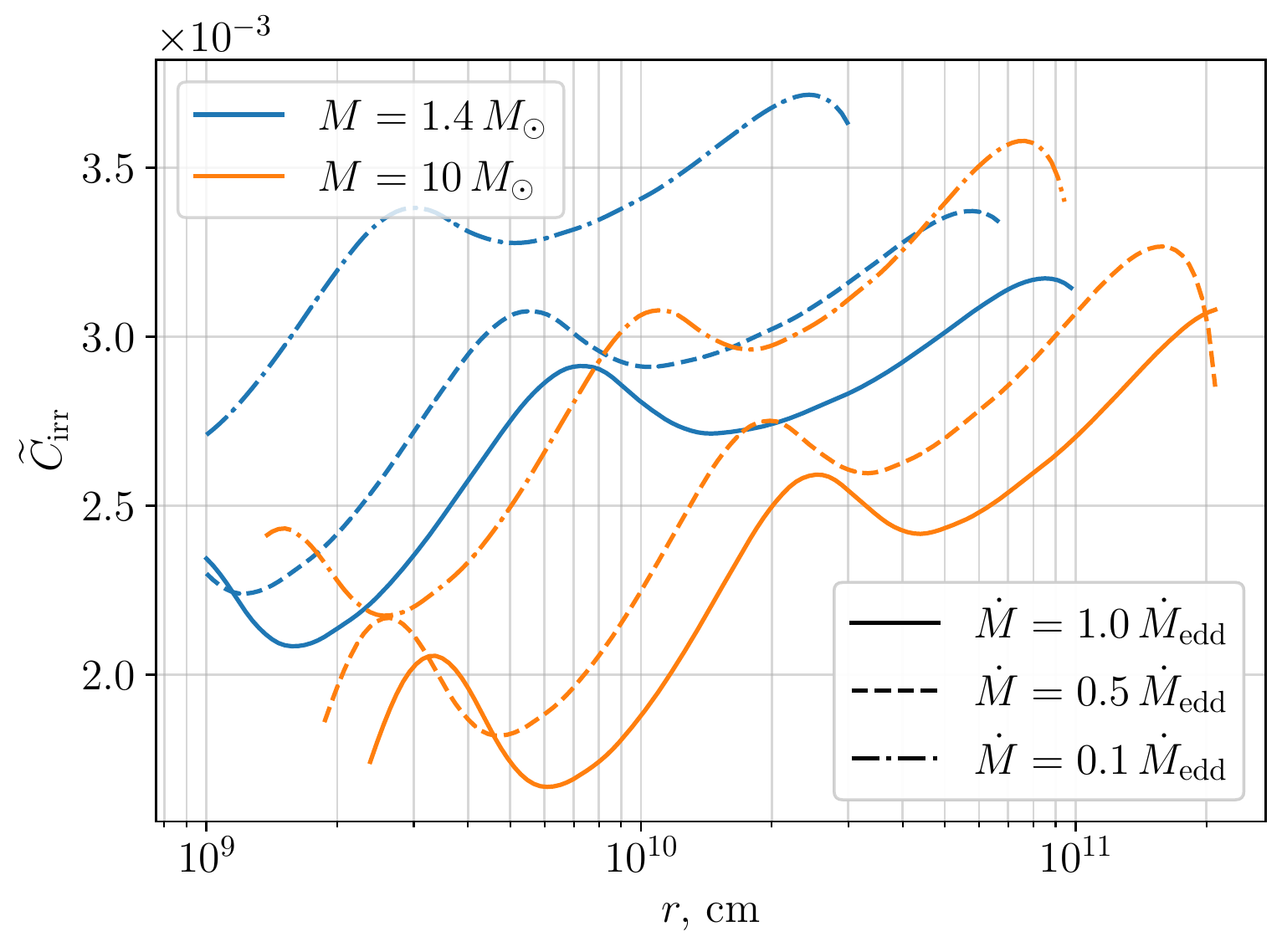}}
	\caption{Radial profile of $\widetilde{C}_{\rm irr}$, system parameters and notations are the same as in Fig.~\ref{fig:C_irr}. It is clearly seen, that this value almost does not change over radius (it changes by a factor of ${\sim}1.5$ by two orders of magnitude along the radius).}
	\label{fig:C_irr_z0r}
\end{figure}

\section{Brief code description}
\label{appendix:Brief-Code-description}

The \texttt{Python~3} code solves the vertical structure equations. It contains several classes which represent the vertical structure for different types of EoS and opacity, temperature gradient and irradiation scheme. Function \texttt{StructureChoice()} serves as an interface to initialize the chosen structure type. The code also contains three functions to calculate S-curves, vertical and radial profile of a stationary disc.

Main input parameters of the code are: mass of the central object~$M$, radius~$r$, viscous torque~$F$ (or effective temperature $T_{\rm eff}$ or accretion rate $\dot{M}$), turbulence parameter~$\alpha$ and type of the structure (depending on the opacity law, EoS, irradiation scheme, temperature gradient, see the GitHub documentation and the code~\texttt{help()}).Notice that different chemical composition can be set if tabular opacity and EoS are used (see Sect.~\ref{subsec:S-types}). 

If irradiation takes place, there are two cases:
\begin{enumerate}
    \item If irradiation is described in terms of $T_{\rm irr}$ or $C_{\rm irr}$, then the code has one additional input parameter~-- irradiation temperature $T_{\rm irr}$ or irradiation parameter $C_{\rm irr}$.
    \item If irradiation is described through the advanced scheme, the external irradiation flux is given by Eq.~(\ref{eq:X-ray-flux}), and the code has a few more input parameters: the X-ray luminosity of central source $L_{\rm X}$; the spectrum $S(\nu)$ in form of table values (normalized over the frequency range to unity) or as a \texttt{Python} function; cosine of the incident angle $\cos\theta_0$ as a fixed value or as a fixed value in the brackets in function~(see also Eq.~(\ref{eq:cos_theta_irr})):
    \begin{equation}
        \cos\theta_0 = \frac{z_0}{r} \left(\frac{{\rm d}\ln z_0}{{\rm d}\ln r}-1\right).
    \end{equation}
\end{enumerate}

Note that the calculated vertical structure of a disc ring without irradiation differs to minor extent when we use alternative boundary conditions on pressure: (\ref{eq:P_bound_full})~and~(\ref{eq:P_irr_bound}). The first boundary condition is implemented as described in section~\ref{subsec:Basic-equations}. The second variant can be engaged in irradiated-disc scheme~(ii) with $C_{\rm irr}=0$ or $T_{\rm irr}=0$.

The free parameter $z_0$ is found using so-called shooting method. The system is integrated with different values of the free parameter, starting from initial estimation, in order to satisfy the additional condition for flux~(\ref{eq:Q_bound_another}) at the symmetry plane of the disc. In the presence of external irradiation in scheme~(i), the only change is the boundary condition for temperature~(\ref{eq:T_irr_bound}). If irradiation is taken into account through the advanced scheme~(ii), the system is modified as described in Sect.~\ref{subsec:Irr_ii} and two-parameter $(z_0, \Sigma_0)$ optimization problem is solved in order to satisfy both the additional boundary conditions~(\ref{eq:Q_bound_another}) and~(\ref{eq:Sigma_bound_another}).

Code is open-source and available with detailed documentation on GitHub\footnote{\url{https://github.com/AndreyTavleev/DiscVerSt}}. \texttt{Scipy}~\citep{Scipy}, \texttt{Numpy}~\citep{Numpy}, \texttt{Matplotlib}~\citep{Matplotlib} and \texttt{Astropy}~\citep{Astropy1, Astropy2} packages are used in the code.

\section{Vertical structure: examples}
\label{appendix:Vertical-structure}

\begin{figure*}
    \center{\includegraphics[width=0.9\linewidth]{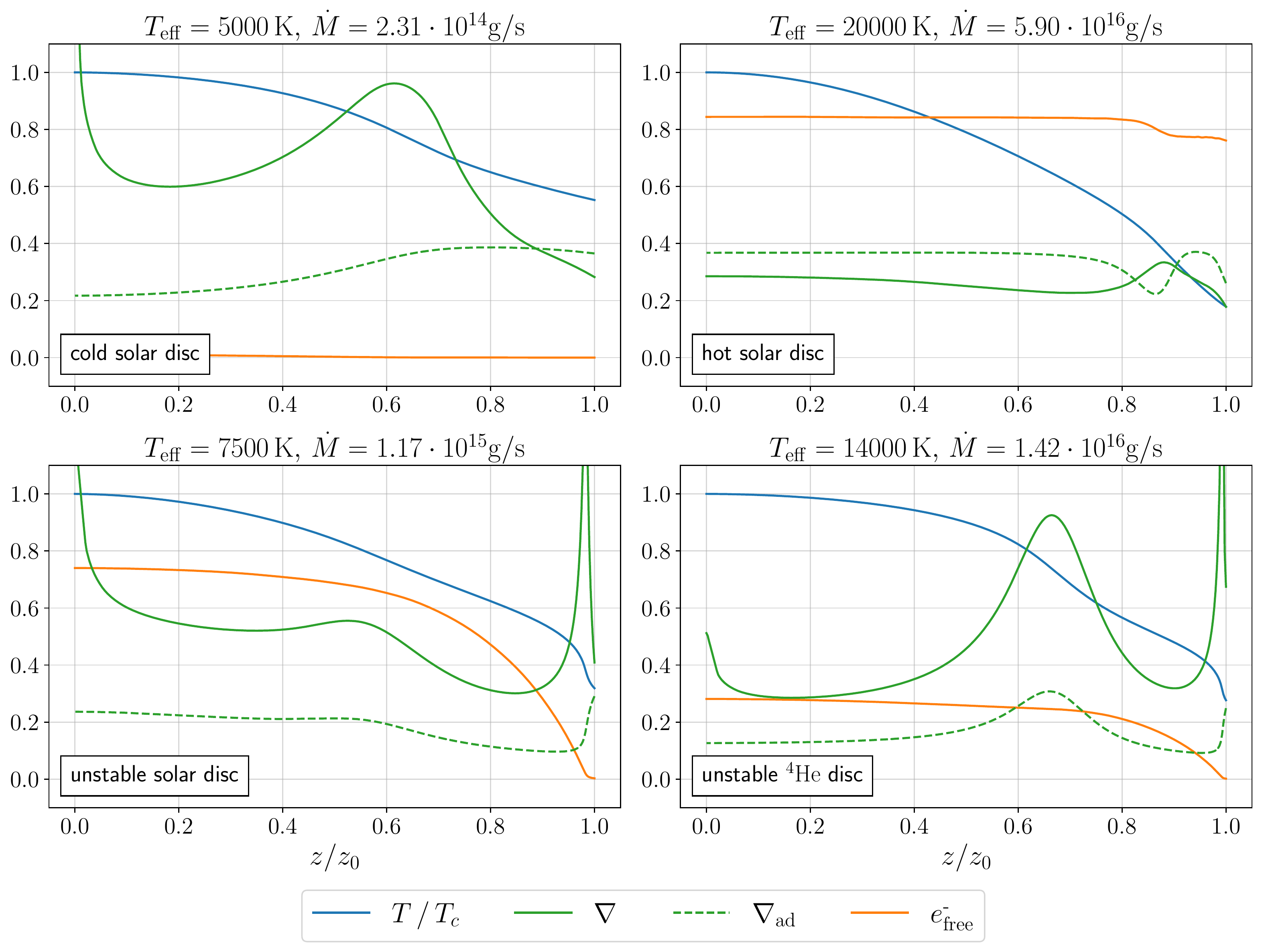}}
    \caption{Disc vertical structure for $M = 10\, M_{\sun}, \alpha = 0.1,\, r = 10^{10}\, \rm cm$ and tabular opacity for different accretion rates~$\dot{M}$ and effective temperatures~$T_{\rm eff}$. Shown are normalized temperature $T / T_c$, actual temperature gradient $\nabla$, adiabatic gradient $\nabla_{\rm ad}$, and mean number of free electrons per nucleon. Upper panels are calculated for cold ($T_{\rm eff} = 5000\, \rm K$) and hot ($T_{\rm eff} = 20000\, \rm K$) disc states with solar composition. Lower panels are calculated for unstable disc states with solar and helium composition. There is no convection in the hot disc ($\nabla_{\rm rad}<\nabla_{\rm ad}$), while the disc in cold neutral and unstable states is convective.}
    \label{fig:vs}
\end{figure*}

\autoref{fig:vs} presents examples of the vertical structure for different effective temperatures, which are determined by the accretion rate, at fixed radius $r=10^{10}\rm\, cm$ for a case without external irradiation. Shown are the temperature distribution, adiabatic and actual temperature gradients, and the mean number of free electrons per nucleon $e_{\rm free}^\text{-} \equiv 1/\mu_{e}$. The latter can change from $0$ in neutral matter to $(1+X)/2$ in fully ionized matter, where $X$ is the hydrogen abundance. 

The upper panels of Fig.~\ref{fig:vs} represent stable disc (in hot and cold state), while lower panels show unstable disc with different chemical composition (solar and pure helium). The latter solutions lie on the negative branch on the S-curve, see Sect.~\ref{sec:S-curves} and Fig.~\ref{fig:S-curve},~\ref{fig:Experiment}. The unstable state is related to ionization of hydrogen: while the cold disc is neutral ($e_{\rm free}^\text{-}\approx0$) and hot disc is fully ionized ($e_{\rm free}^\text{-}\approx0.85$), ionization of unstable disc changes along $z$ between these two limits.

The disc in cold and unstable state is convective ($\nabla_{\rm rad}>\nabla_{\rm ad}$ along the $z$ coordinate), while there is no convection in the hot disc (except for a thin layer near the surface). This happens regardless of the chemical composition: the pure helium disc behaves similarly. The main difference is that instability in helium disc is related to the partial ionization of helium, therefore the temperature of unstable disc ($T_{\rm eff}\sim 14\,000 \,\rm K$) is higher than in that in solar disc ($T_{\rm eff}\sim 7000 \,\rm K$). The corresponding S-curves are presented in Sect.~\ref{subsec:S-types}, see Fig.~\ref{fig:Experiment}.

\begin{figure*}
	\center{\includegraphics[width=1.0\linewidth]{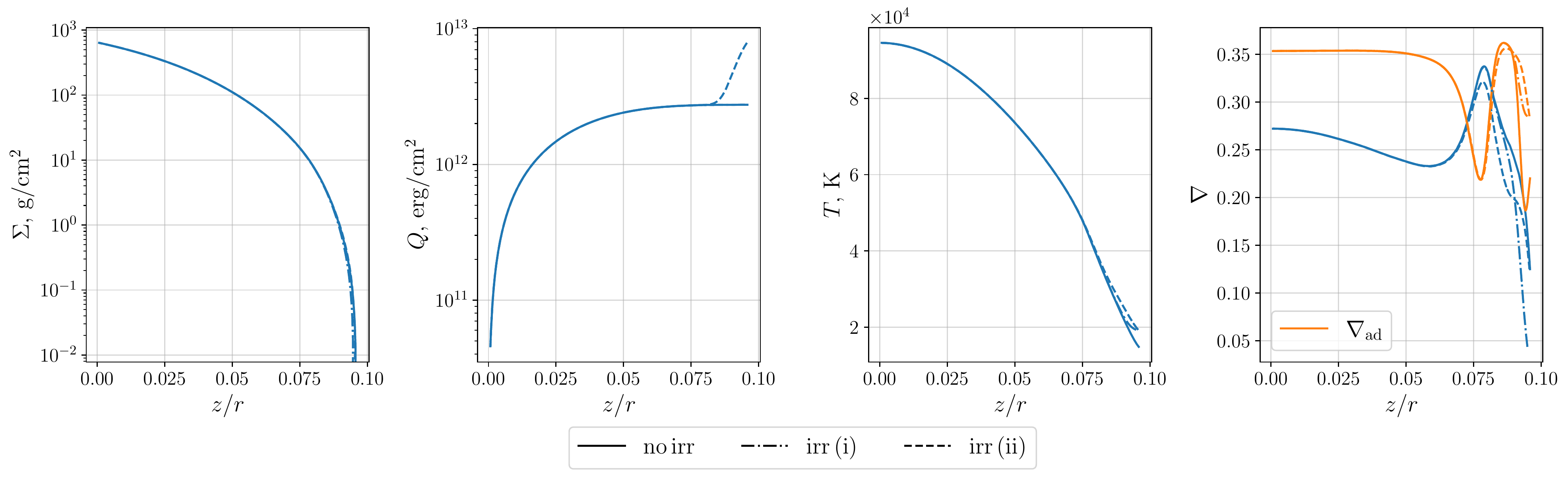}}
	\caption{Vertical structure of irradiated disc together with un-irradiated case for $M = 1.4\, M_{\sun}, \,\alpha = 0.1,\, r = 2\cdot 10^{10}\, \rm cm$ and tabular opacity for accretion rate~$\dot{M}=10^{18}\,\rm g\, s^{-1}$. Shown are mass coordinate $\Sigma$, temperature $T$, temperature gradient $\nabla$, adiabatic gradient $\nabla_{\rm ad}$, and energy flux $Q$. Irradiation is taken into account through two approaches~(i) and~(ii), where $L_{\rm X}=\eta\dot{M} c^2$. Irradiation temperature $T_{\rm irr}=17330\,\rm K$ is obtained from calculations by method~(ii) and serves as input parameter in method~(i). The corresponding $C_{\rm irr} = 2.84\cdot 10^{-4}$.}
	\label{fig:vs_irr_2e10}
\end{figure*}

\begin{figure*}
	\center{\includegraphics[width=1.0\linewidth]{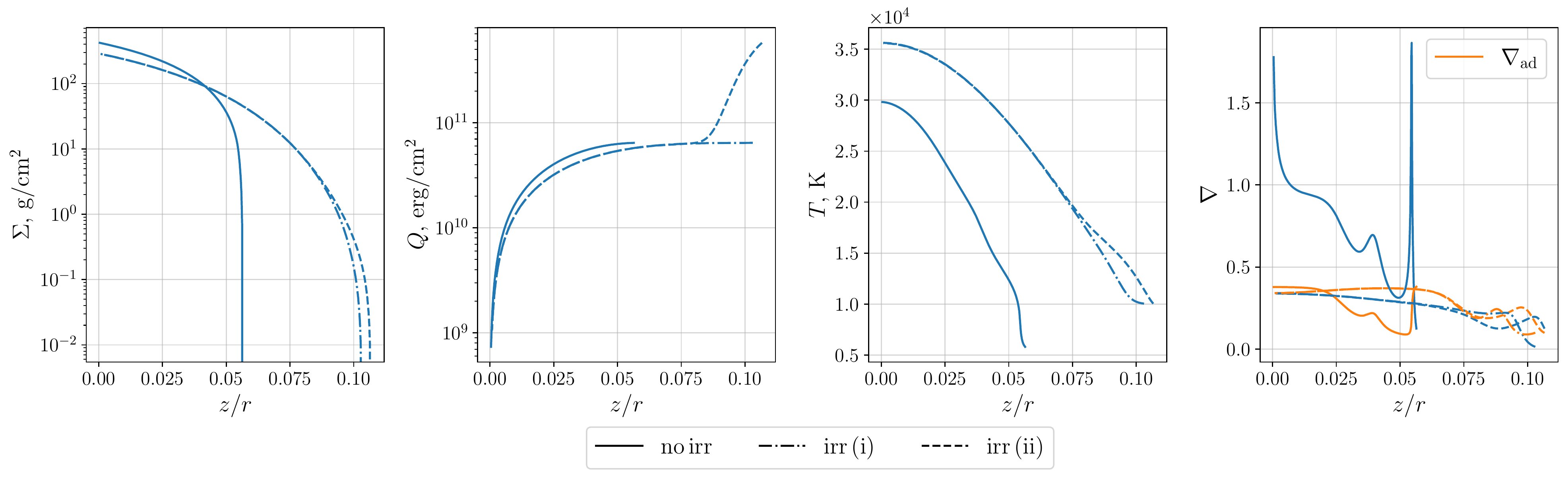}}
	\caption{Vertical structure of irradiated disc together with un-irradiated case. System parameters and notations are the same as in Fig.~\ref{fig:vs_irr_2e10}, but the radius $r=7\cdot 10^{10}\,\rm cm$. The obtained $T_{\rm irr} = 9750 \,\rm K$, and corresponding $C_{\rm irr} = 3.5\cdot 10^{-4}$.}
	\label{fig:vs_irr_7e10}
\end{figure*}

Figures~\ref{fig:vs_irr_2e10} and~\ref{fig:vs_irr_7e10} show examples of the vertical structure for irradiated disc together with un-irradiated case at two radii $r=2\cdot 10^{10} \text{ and } 7\cdot 10^{10}\rm\, cm$. Other system parameters are the same as in Fig.~\ref{fig:r_plots_irr}: $M = 1.4\, M_{\sun}, \,\alpha=0.1, \,L_{\rm X}=\eta\dot{M}c^2, \,\eta=0.1,\,\dot{M} = 10^{18} \, {\rm g\, s^{-1}}\approx 0.5\,\dot{M}_{\rm edd}$, the chemical composition is solar. Shown are distributions of mass coordinate $\Sigma$, temperature $T$, flux $Q$, temperature gradient $\nabla$ and adiabatic temperature gradient $\nabla_{\rm ad}$. 
Irradiation is taken into account through two methods~(i) and~(ii), see Sect.~\ref{subsec:Irradiation-by-central-X-ray-source}. Irradiation temperature $T_{\rm irr}$ is obtained in advanced method~(ii) and serves as input parameter in method~(i). 

It is clearly seen that at $r=2\cdot 10^{10} \rm\, cm$ the external irradiation with $T_{\rm irr} = 17330 \,{\rm K}$ almost does not affect the structure of the disc, whose viscous flux corresponds to $T_{\rm vis}=14840\, {\rm K}$. The irradiation does not penetrate deep into the disc and heats only the near-surface layers, which is seen on the flux dependence in Fig.~\ref{fig:vs_irr_2e10}. Energy in the disc is transferred mainly by radiation (see the upper right panel).

At a larger radius $r = 7\cdot 10^{10}\rm\, cm$, the un-irradiated disc with $T_{\rm vis} = 5800\, {\rm K}$ is unstable and fully convective, but irradiation with $T_{\rm irr} = 9750 \,{\rm K} $, affecting the whole disc in the vertical direction, stabilizes the disc structure at a lower surface density and a larger thickness. Without irradiation, opacity $\varkappa_{\rm R}$ drops in the upper layers by more than two orders. With irradiation, opacity changes with $z$ not more than few times. The convection disappears in the irradiation-stabilized disc.


\bsp	
\label{lastpage}
\end{document}